\documentclass[twocolumn,showpacs,aps,prd]{revtex4}

\usepackage{graphicx}
\usepackage{amsmath,amssymb}
\usepackage{pstricks}
\usepackage{pspicture}
\usepackage{pst-plot}

\begin{document}

\def\cs#1#2{#1_{\!{}_#2}}
\def\css#1#2#3{#1^{#2}_{\!{}_#3}}

\title{Passage of Time in a Planck Scale Rooted Local Inertial Structure}

\author{Joy Christian}

\email{joy.christian@wolfson.oxford.ac.uk}

\affiliation{Wolfson College, Oxford University, Oxford OX2 6UD, United Kingdom}

\thanks{Journal Reference: Int. J. Mod. Phys. D {\bf 13} (2004) 1037-1071}

\date{10 August 2003}

\begin{abstract}
It is argued that the `problem of time' in quantum gravity necessitates a refinement of the
local inertial structure of the world, demanding a replacement of the usual Minkowski line
element by a ${(4+2n)}$-dimensional pseudo-Euclidean line element, with the extra $2n$ being the
number of internal phase space dimensions of the observed system. In the refined structure,
the inverse of the Planck time takes over the role of observer-independent conversion factor
usually played by the speed of light, which now emerges as an invariant but derivative quantity.
In the relativistic theory based on the refined structure, energies and momenta turn out to be
invariantly bounded from above, and lengths and durations similarly bounded from below, by their
respective Planck scale values. Along the external timelike world-lines, the theory naturally
captures the `flow of time' as a genuinely structural attribute of the world. The theory also
predicts expected deviations---suppressed quadratically by the Planck energy---from the
dispersion relations for free fields in the vacuum. The deviations from the special relativistic
Doppler shifts predicted by the theory are also suppressed quadratically by the Planck
energy. Nonetheless, in order to estimate the precision required to distinguish the theory
from special relativity, an experiment with a binary pulsar emitting TeV range ${\gamma}$-rays
is considered in the context of the predicted deviations from the second-order shifts.
\end{abstract}


\maketitle

\section{INTRODUCTION}

Soon after the final formulation of general relativity, Einstein argued that
quantum effects must modify the new theory of gravity \cite{Einstein-1916}.
Despite almost a century of debate, however, today there is little consensus on
what eventual form such a modified theory should take \cite{Christian-2001}. One
of the few notions that does enjoy unanimity in this context is the recurrent suggestion
of a `minimum length', which is usually taken to be the Planck length
${{\cs lP}:=\sqrt{\hbar\,{\cs GN}/c^3}}$. It is generally believed that the Planck scale
(cf. p. 471 of \cite{Wald-1984}) marks a threshold beyond which the usual continuum
description of spacetime is unlikely to survive. Already in the wake of Einstein's
pioneering argument there existed a number of speculative proposals for discrete spacetimes.
For instance, Heisenberg suggested that the spatial continuum should be replaced by
a `lattice world', composed of cells of a finite size \cite{Heisenberg-1938,Heisenberg-1930}.
Later he abandoned the idea, however, concluding that the assumption of minimum length
appeared to be incompatible with Lorentz invariance. Indeed, an elementary implication of
Lorentz contraction reveals that an assumed minimum length in one inertial frame may have
a different---and even vanishing---value in another.

It is now well known that in a remarkable paper Snyder \cite{Snyder-1947} 
eventually resolved this apparent
incompatibility between the existence of a minimum length and Lorentz invariance
(see also \cite{Yang-1947}). In the modern parlance,
Snyder's resolution amounts to finding a {\it non-linear} basis for the Poincar\'e
algebra that allow {\it two}, instead of one, observer-independent scales; namely, the
maximum speed ${c}$ as well as a minimum length ${\cs lP}$. Recent years have seen a
revival of theories allowing two observer-independent scales based on such non-linear actions
of the Poincar\'e group \cite{Amelino-Camelia-2001, Magueijo+Smolin-2002,
Kowalski-Glikman-2002, Lukierski+Nowicki-2003}. Physically, these theories 
correspond to non-linear modifications of special relativity, allowing the low-energy photons
in some cases to travel faster than the upper bound ${c}$. One may classify these neo-Snyder
theories as `bottom-up' attempts to construct quantum gravity, in a manner not too dissimilar
from how general relativity was founded by Einstein on special relativity.
 
The theory developed in this paper shares the general `bottom-up' philosophy of these
approaches. However, it also fundamentally differs from them in a number of different
ways. For instance, as expected of any framework underlying quantum gravity, the {\it only}
fundamental scale in the theory proposed here is the Planck scale; although the vacuum speed
of light remains an observer-independent absolute upper bound on speeds as before, it now plays
only a secondary and derivative role. What is more, unlike the diversity of the two-scales
theories, the theory developed here is {\it unique}, with energies and momenta invariantly
bounded from above, and lengths and durations similarly bounded from below, by their respective
Planck scale values. Away from the Planck scale the theory reduces in general, as any such
theory must, to Einstein's special theory of relativity.

Perhaps most significantly, and at least compared to the two-scales theories (cf.
\cite{Ahluwalia++, Ellis-2003}), the theory proposed here is based on operationally better
grounds, thereby retaining the original rationale of Einstein's one-scale theory. In
particular, the central aim of the proposed theory is to eliminate the operationally
questionable dualistic conception of time implicitly taken for granted in physical theories.
This leads to a {\it generalization} of Lorentz invariance that is fundamentally different form
either the conception of a `broken Lorentz invariance'---as in some approaches to quantum
gravity \cite{Kostelecky-1989}, which may give rise to a `preferred frame'---or the conception
of `modified Lorentz invariance'---as in the two-scales theories \cite{Magueijo+Smolin-2002},
which may not give rise to a `preferred frame'. By contrast, the generalized invariance
employed in the present theory is---by its very essence---hostile to the notion of a `preferred
frame'. In fact, the employed invariance respects the principle of relativity fully across the
conceptual and physical domains that are even broader than those in Einstein's theories of
relativity.

As expected within at least some of the mainstream approaches to quantum gravity (for a review,
see \cite{Smolin-2003}), the theory proposed here also predicts (exact) deviations---suppressed
{\it quadratically} by the Planck energy---from the dispersion relations for the propagation of
free fields in the vacuum. In recent years it has become increasingly probable that theories
predicting even such minute deviations from the dispersion relations may be subjected to
observational tests (see \cite{Smolin-2003, Alfaro-2003, Amelino-Camelia-2003, Sarkar-2002,
Jacobson-2003}
and references therein). In addition to these possibilities, we shall also attempt to estimate
the precision required to distinguish the modified Doppler shifts predicted by the present
theory from those predicted by special relativity, by considering an experiment with a binary
pulsar emitting TeV range ${\gamma}$-rays.

One of the oldest issues in natural philosophy is `the problem of change' \cite{Popper}. Since
the days of Aristotle, physics has been tremendously successful in explaining {\it how} the
changes occur in the world, but largely oblivious to the deeper question of {\it why} do they
occur at all. The situation has been aggravated by the advent of Einstein's theories of
spacetime, since in these theories there is no room to accommodate a {\it structural}
distinction between the past and the future \cite{Davies-1974}---a prerequisite for a true
understanding of {\it why} the changes occur in the world. By contrast, the causal structure of
the proposed theory  below not only naturally distinguishes the future form the past, but also
{\it forbids} inaction altogether, thereby providing an answer to the deeper question of change.

In what follows we begin with the conceptual basis of the proposed theory in Secs. II to V,
culminating in a ${(4+2n)}$-dimensional fundamental quadratic invariant, namely (\ref{quad}).
The basic physical consequences of the theory erected on this new quadratic invariant are then
spelt out in Sec. VI. In the following section, Sec. VII, we raise and answer the age-old
question: `How fast does time flow?' (see, e.g., \cite{Lockwood-1989}). Since the proposed
theory naturally quantifies the motion of the present moment (or `now') along timelike
world-lines of observers, what we have is an experimentally verifiable complete theory of the
local inertial structure that captures the `flow of time' as a genuine attribute of the world.
Finally, before concluding in Sec. IX, the options for possible experimental
verifiability of the theory are explored in Sec.VIII.

\section{Time in Special Relativity}

Recall that among Einstein's primary concerns while constructing special relativity were the
notions of absolute time and relative motion. In particular, he noted that some of the
assumptions underlying these concepts within the then existing framework of physics were
operationally ill-founded. To rectify this inadequacy, he proposed an operationally better
founded structure we now take to be the local inertial structure of the world. As is well known,
this structure---later geometrically refined by Minkowski---is based on
the following two postulates:
(i) {\it The laws governing the states of physical systems are insensitive to the
state of motion of the reference coordinate system, as long as it remains inertial},
and (ii) {\it No speed of a causally admissible influence can exceed the vacuum speed of light}.
The second of the two postulates can be succinctly restated as
\begin{equation}
v\leq c\,,\label{c-bound}
\end{equation}
with the understanding that it is a statement true for all local inertial
observers regardless of their state of motion. This is reaffirmed by the
law of composition of velocities proposed by Einstein: assuming all velocities
involved have the same direction, the velocity of a material body, say ${v_{}^k}$
(${k=1,2,}$ or ${3}$),
in one inertial frame is related to its velocity, say ${v'^k}$, in another
frame, moving with a velocity ${-v^k_r}$ with respect to the first, by the relation
\begin{equation}
v'^k=\frac{v_{}^k+\,v^k_r}{1+c^{-2}\,v_{}^k\,v^k_r}\,.\label{v-relation}
\end{equation}
Thus, as long as neither ${v_{}^k}$ nor ${v^k_r}$ exceeds the causal upper bound
${c}$, ${v'^k}$ also remains within ${c}$. It is this absoluteness of ${c}$
that lends credence to the view that it is merely a conversion factor between
the dimensions of time and space. This fact is captured most conspicuously by
the four-dimensional quadratic invariant of spacetime,
\begin{equation}
d{\css t2E}:=d{\css t2N} - c^{-2}d{\bf x}^2\,,\label{e-metric}
\end{equation}
where ${\bf x}$ is a vector in the 3-space, and ${\cs tN}$ and
${\cs tE}$ are the Newtonian (or absolute or coordinate) and Einsteinian (or proper
or wristwatch)
times, respectively. In natural units where ${c}$ is taken to be unity, this line element can
be put in the familiar Minkowski form:
\begin{equation}
d{\css t2E}:=d{\css t2N} - d{\bf x}^2=:-\eta_{ab}\,dx^adx^b\,.\label{e-natural}
\end{equation}
The central tenet of special relativity is now simply the
assertion that it is the time interval
${d{\cs tE}}$, and not the Newtonian interval ${d{\cs tN}}$,
that is actually registered by an inertial
observer between two nearby events in spacetime, with
the causality restriction equivalent to (\ref{c-bound}) being
\begin{equation}
d{\css t2E}\geq 0.\label{causality}
\end{equation}
To date there exists no
convincing experimental evidence contradicting the varied predictions of
Einstein's theory.

Despite the phenomenal empirical successes of special relativity, however,
a closer inspection reveals that, as yet, Einstein's first postulate---the
principle of relativity ---is not fully respected within the current framework
of physics. As we shall see, when thoroughly implemented, the principle
necessitates generalizations of the upper bound (\ref{c-bound}) and the
law of composition (\ref{v-relation}), which in turn demand a Planck scale
amelioration of the metric (\ref{e-metric}).

\section{Time in quantum gravity}

To appreciate the above assertions, consider an object system, equipped with an ideal
classical clock of unlimited accuracy, moving with a uniform velocity ${\bf{v}}$ in a
Minkowski spacetime ${\cal M}$, from an event ${{\rm e}_1}$ at the origin of a reference frame
to a nearby event ${{\rm e}_2}$ in the future light cone of ${{\rm e}_1}$. For our purposes, it
would suffice to refer to this system, say of ${n}$ degrees of freedom, simply as `the clock'
(cf. \cite{Wigner-1958}). As it moves, the clock will also necessarily {\it evolve} by the
virtue of its `external' motion, say at a uniform rate ${\boldsymbol{\omega}}$, from one
state, say ${{\rm s}_1}$, to another, say ${{\rm s}_2}$, within its own `internal' relativistic
phase space, say ${\cal N}$. This evolution of the clock---or rather that of its state---from
${{\rm s}_1}$ to ${{\rm s}_2}$ will, of course, trace out a unique trajectory in the phase
space ${\cal N}$, which, as is customary for any state space, may be parameterized by the
Newtonian time ${\cs tN}$ and viewed as a curve in the resulting ${(1+2n)}$-dimensional extended
phase space (or contact manifold) \cite{Arnold-1989}. For simplicity, here we shall
assume the phase space to be finite and ${2n}$ dimensional; apart from possible mathematical
encumbrances, however, nothing prevents the following reasoning to remain valid for the
case of infinite dimensional phase spaces (e.g. for clocks made out of relativistic fields).

Now, one may think of the motion and evolution of the clock conjointly as taking place
in a combined ${(4+2n)}$-dimensional space, say ${\cal E}$, the elements of which may be
called {\it event-states} and represented by pairs ${({\rm e}_i,\,{\rm s}_i)}$.
Undoubtedly, it is this combined space that truly captures the complete specification
of all possible physical attributes of the classical clock. Therefore, we may ask:
{\it What will be the time interval actually registered by the clock as it
moves and evolves from the event-state ${({\rm e}_1,\,{\rm s}_1)}$ to the event-state
${({\rm e}_2,\,{\rm s}_2)}$ in this combined space ${\cal E}$}? It is
only by answering such a physical question that one can determine the correct topology and
geometry of the combined space in the form of a metric analogous to (\ref{e-metric}).

Of course, the customary answer to the above question, in accordance with the
spacetime metric (\ref{e-metric}), is simply
\begin{equation}
\Delta {\cs tE}=\int_{e_1}^{e_2}\frac{1}{\gamma(v)}\;d{\cs tN}
=\frac{\left|t_2-t_1\right|}{\gamma(v)}\,,\label{not-this}
\end{equation}
with the usual Lorentz factor
\begin{equation}
1<\gamma(v):=\left(1-c^{-2}v^2\right)^{-\frac{1}{2}}.\label{x-gamma}
\end{equation}
In other words, the customary answer is that the space ${\cal E}$ has a product topology,
${{\cal E}={\cal M}\times{\cal N}}$, and---more to the point---the clock recording the
duration ${\Delta{\cs tE}}$ in question is {\it insensitive} to the passage of time that
marks the evolution of variables within its own phase space ${\cal N}$.

This may appear to be a perfectly adequate answer. After all, a classical clock is
supposed to register the time {\it external} to itself, and the phase space of its evolution
is an {\it internal} space, dependent on the physical attributes of the clock. In other
words, the time measured by such a clock is presumed to be `external', existing independently
of it, whereas its dynamic evolution, although parameterizable by this background time, is
viewed to
be `internal', specific to the clock itself. Such a dualistic conception of time has
served us well in special relativistic physics. In fact, we encounter no serious difficulties
with its implementation in physical theories---classical or quantal---until quantum gravity is
confronted head-on \cite{Penrose-1996}.

The formidable difficulties in constructing quantum gravity, of course, stem from the fact
that the basic principles of general relativity and quantum mechanics are fundamentally
at odds with each other \cite{Christian-2001}. One of the many ways these difficulties can
manifest themselves is via the so-called `problem of time' within the canonical approaches
to the task \cite{Isham-1993}. The problem, essentially, is that, unlike in special and
general theories of relativity, in classical and quantum mechanics time is treated as an
external parameter, marking the evolution of a given physical system. Within the above
approaches, most attempts that try to reconcile the dynamical role of time in
relativity with the parametric role of time in mechanics seek to understand this external
parameter in terms of some internal structure of the system itself \cite{Isham-1993}.

Needless to say, to date no such attempt has succeeded in providing unequivocal
understanding of the `external' time that observers do register on their clocks. What is worse,
the dichotomy of time into external versus internal time emphasized by most such approaches to
quantum gravity only manages to obscure the real culprit lurking behind the difficulty;
namely, the dichotomy of {\it simultaneity} into {\it relative} versus {\it absolute}
simultaneity. Unlike in spacetime physics, where the unphysical notion of absolute simultaneity
was eliminated by Einstein, in {\it state}-time physics, the notion remains axiomatic. It is
in this sense, then, that Einstein's first basic postulate---the principle of
relativity---is not as yet fully respected within the established framework of physics.

\section{proper Time in the `rest Frame'}

Let us look at these assertions more closely by unpacking them once again with the help of the
moving and evolving clock considered above. In the `global inertial frame' \cite{Wald-1984}
in which the clock is at rest, the proper time interval it is supposed to register is simply
${d{\cs tE}=d{\cs tN}}$. Using this Newtonian time ${\cs tN}$ as an external parameter, within
this frame one can determine the phase space ${\cal N}$ for the dynamical evolution of the clock
using the standard Hamiltonian prescription \cite{Goldstein-1980}. Of course, as long as
we start with a Lorentz invariant action, the resulting phase space within each such
frame---although not manifestly covariant---would be consistent with the principles of special
relativity. Suppose now we consider time-dependent canonical transformations of the
dimensionless phase space coordinates ${{\rm y}^{\mu}({\cs tN})}$
(${\mu=1,\ldots,2n}$), expressed in Planck units, into coordinates
${{\rm y}'^{\mu}({\cs tN})}$ of the following general linear form:
\begin{equation}
{\rm y}'^{\mu}({\rm y}^{\mu}(0),\,{\cs tN})={\rm y}^{\mu}(0)+
\omega_r^{\mu}({\bf y}(0))\,{\cs tN}+b^{\mu}\,,\label{cano-trans}
\end{equation}
where ${\omega_r^{\mu}}$ and ${b^{\mu}}$ do not have explicit time dependence, and the
reason for the subscript ${r}$ in ${\omega_r^{\mu}}$, which stands for `relative', will
become clear soon.
Interpreted actively, these are simply the linearized solutions of the familiar
Hamiltonian flow equations \cite{Arnold-1989},
\begin{equation}
\frac{d{\rm y}^{\mu}}{d{\cs tN}}=\omega_r^{\mu}\left({\bf y}({\cs tN})\right)
:=\Omega^{\mu\nu}\frac{\partial H}{\partial {\rm y}^{\nu}}\,,\label{H-eqn}
\end{equation}
where ${{\boldsymbol\omega}_r}$ is the Hamiltonian vector field generating the flow,
${{\bf y}({\cs tN})}$ is a ${2n}$-dimensional local Darboux vector in the phase space
${\cal N}$, ${\boldsymbol{\Omega}}$ is the symplectic 2-form on ${\cal N}$, and ${H}$ is a
Hamiltonian function governing the evolution of the clock. In the dual description, the vector
field ${{\boldsymbol\omega}_r}$ generating the integral curves of (\ref{H-eqn})
can be defined by
\begin{equation}
\omega_{\mu}^r{\bf d}{\rm y}^{\mu}:=
{\boldsymbol{\Omega}}\left(\;\cdot\;,\,{{\boldsymbol\omega}_r}\right)={\bf d}H
\equiv\frac{\partial H}{\partial {\rm y}^{\mu}}{\bf d}{\rm y}^{\mu},\label{one-form}
\end{equation}
with ${\bf d}$ being the exterior derivative. Since
${{\boldsymbol{\Omega}}\left(\;\cdot\;,\,{{\boldsymbol\omega}_r}\right)}$---with only one of
the two slots of the 2-form ${\boldsymbol{\Omega}}$ filled---is a 1-form dual to the vector
field ${{\boldsymbol{\omega}}_r\,}$,
its components are denoted by ${\omega_{\mu}^r}$. In other words,
as in any phase space, the indices ${\mu}$ of the vector field ${\omega_r^{\mu}}$ is lowered,
not by a possible metric on ${\cal N}$, but by the symplectic 2-form ${\boldsymbol{\Omega}}$.
Being antisymmetric, however, the 2-form ${\boldsymbol{\Omega}}$ is not suitable as a
metric on ${\cal N}$. Fortunately, at least locally, one can define a Euclidean metric
${\boldsymbol{\delta}}$ on ${\cal N}$ to evaluate inner products. Using this metric,
it is then easy to see from relations (\ref{H-eqn}) and (\ref{one-form}) that, when
Hamiltonian happens to be the total energy of the system, the dimensionless
magnitude of the
vector field ${{\boldsymbol\omega}_r}$ is given by
\begin{equation}
{\cs tP}\,\omega_r=\frac{d E}{{\cs EP}}\,,\label{Hamilton-Jacobi}
\end{equation}
where ${{\cs EP}}$ is the Planck energy, which, along with ${{\cs tP}}$, explicates the
balance of dimensions on the two sides.

For a later use, let us make a few comments on the physical meaning of the
Euclidean metric ${\boldsymbol{\delta}}$ on the phase space and the corresponding
dimensionless line element
\begin{equation}
d{\bf y}^2\!:=\delta_{\mu\nu}\,d{\rm y}^{\mu}d{\rm y}^{\nu}.\label{y-metric}
\end{equation}
In general, every symplectic manifold ${({\cal N},\,{\boldsymbol{\Omega}})}$ admits a family of
compatible almost complex structures \cite{Analysis}, but none of these need be integrable. When
one is, it gives rise to a Riemannian metric on ${({\cal N},\,{\boldsymbol{\Omega}})}$ defined
by ${\bf g(X,\,Y):={\boldsymbol{\Omega}}(X,\,JY)}$, where ${{\bf X},{\bf Y}}$ are two arbitrary
vector fields and ${\bf J}$ is the complex structure. A prime example of such a complex manifold
(known as a K\"ahler manifold) is the projective Hilbert space, with its well known Fubini-Study
metric of a constant holomorphic curvature. Now, of course, this Fubini-Study metric gives rise
to transition probabilities in quantum mechanics and reduces to the Euclidean metric
${\boldsymbol{\delta}}$ in the classical limit, whereas the corresponding
symplectic form gives rise to the celebrated geometric phase factor and reduces
to the classical symplectic form ${\boldsymbol{\Omega}}$ in the classical
limit\cite{Anandan-1990}. Therefore, the `flat' Euclidean metric ${\boldsymbol{\delta}}$
(or the element (\ref{y-metric})) of our phase space can be viewed as the `quantum shadow' of
the `curved' quantum state space metric (i.e., of the Fubini-Study metric), also giving rise to
transition probabilities---albeit
of a rather trivial kind, yielding only extreme values of 0 or 1 (cf. \cite{Fivel-1994}).

Now, since a phase space in general is simply a bare symplectic manifold, one may wonder about
those clocks whose mathematical phase space descriptions do not require (or even admit) such a
`quantum shadow'. Since the `non-metrical' phase space description of such a clock can only be
recovered in a singular unphysical limit (${\hbar=0}$) from the metrical quantum state space
description in terms of the Fubini-Study metric,
in the present theory the pure symplectic description
of a clock will be deemed `too classical', and the corresponding clock will be deemed `too
crude' (i.e., not sensitive enough to resolve the Planck scale effects predicted by the present
theory). Conversely, it has been argued by Klauder that any consistent quantization scheme
inevitably ends up using the Euclidean metric, if only implicitly, and that in general it is
impossible to ascribe any operational meaning to coordinatized phase space expressions for
physical quantities, such as Hamiltonians, if the corresponding phase space does not admit the
Euclidean metric \cite{Klauder-1997, Klauder-2001}. Therefore, in what follows it will be taken
for granted that the phase space of our Planck scale sensitive clock is equipped with the
`quantum shadow metric' ${\boldsymbol{\delta}}$. In other words, in line with the
`bottom-up' philosophy advocated in the Introduction, it will be taken for granted that it is
this remnant of quantum mechanics---this `quantum shadow metric' ${\boldsymbol{\delta}}$---that
renders our clock sensitive to the Planck scale effects predicted by the present theory.

If we now denote by ${\omega^{\mu}}$ the uniform time rate of change of the canonical
coordinates ${{\rm y}^{\mu}}$,
then the time-dependent canonical transformations (\ref{cano-trans}) imply the composition law
\begin{equation}
\omega'^{\mu}=\omega_{}^{\mu}+\omega_r^{\mu}\label{omega-trans}
\end{equation}
for the evolution rates of the two sets of coordinates, with ${-\omega_r^{\mu}}$ interpreted
as the rate of evolution of the transformed coordinates with respect to the original ones
(in fact, Arnold \cite{Arnold-1989}, for example, calls
${{\boldsymbol\omega}_r}$ simply `the phase velocity vector field'). Crucially for our
purposes, what is implicit in the law (\ref{omega-trans}) is the assumption that there is no
upper bound on the
rates of evolution of physical states. Indeed, successive transformations of the type
(\ref{cano-trans}) can be used, along with (\ref{omega-trans}), to generate arbitrarily high
rates of evolution for the state of the clock.

More pertinently, the assumed validity of the composition law (\ref{omega-trans}) of evolution
rates turns out to be equivalent to assuming absolute simultaneity within the
${(1+2n)}$-dimensional extended phase space \cite{Arnold-1989}, say ${\cal O}$. To appreciate
this fact, let us view ${\cal O}$ as an abstract manifold, and compare the causal structure of
its elements---which may be called {\it occasions} ${o_i}$, representing the pairs
${o_i:=(t_i,\,s_i)}$---with the familiar causal structure of events in spacetime
(see, e.g., \cite{Wald-1984}). It is clear that the causal relationships between two such
occasions, say ${o_i}$ and ${o_j}$, fall into the following three mutually exclusive
possibilities: (i) It is possible, in principle, for a state of a physical system to evolve
from occasion ${o_j}$ to occasion ${o_i}$, in which case ${o_j}$ is said to be in the past of
${o_i}$. (ii) It is possible, similarly, for a state to evolve from occasion ${o_i}$ to occasion
${o_j}$, in which case ${o_j}$ is said to be in the future of ${o_i}$. (iii) It is
{\it impossible}, in principle, for a state of a physical system to be at both occasions
${o_i}$ and ${o_j}$. Now, if we further assume the validity of the composition law
(\ref{omega-trans}), or equivalently the possibility of limitlessly high rates of
evolution for physical states, then, in analogy with events in spacetime, the occasions in
${\cal O}$ belonging to the
third category above would form a ${2n}$-dimensional set (i.e. a phase space) defining the
notion of absolute simultaneity with ${o_i}$. Thus in this case, within the
${(1+2n)}$-dimensional manifold ${\cal O}$, the ${2n}$-dimensional phase spaces
simply constitute strata of `hypersurfaces of simultaneity', much like the
strata of spatial hypersurfaces within a Newtonian spacetime. Indeed, the extended phase spaces
such as ${\cal O}$ are usually taken to be contact manifolds \cite{Arnold-1989},
with topology presumed to be a product of the form ${\mathbb{R}\times{\cal N}}$. 

Thus, not surprisingly, the assumption of absolute time in contact spaces, in defiance of the
relativity principle, is equivalent to the assumption of `no upper bound' on the
possible rates of evolution of physical states. Now, as a variant of the recurrent suggestion
of `a minimum length' (or of a `minimum time'; cf. \cite{Iwanenko-1930}) discussed in the
Introduction, suppose we impose the following upper bound on the rates of evolution of the
state of our clock:
\begin{equation}
\left|\frac{d{\bf y}({\cs tN})}{d{\cs tN}}\right| =: \omega\leq {\css t{-1}P}\,,\label{t-bound}
\end{equation}
where ${\cs tP}$ is the Planck time. Clearly, if this upper bound is to have any physical
significance, it must hold for {\it all} possible evolving phase space coordinates
${{\rm y}^{\mu}({\cs tN})}$ discussed above. And that is amenable if and only if the
composition law (\ref{omega-trans}) is replaced by
\begin{equation}
\omega'^{\mu}=\frac{\omega_{}^{\mu}+\,\omega^{\mu}_r}{1+{\css t2P}\;
\omega_{}^{\mu}\,\omega_r^{\mu}}\label{w-relation}
\end{equation}
(no summation over ${\mu}$ --- cf. the composition law (\ref{v-relation})),
which implies that as long as neither ${\omega^{\mu}}$ nor ${\omega^{\mu}_r}$ exceeds the
causal upper bound ${\css t{-1}P}$, ${\omega'^{\mu}}$ also remains within ${\css t{-1}P}$.

Along with (\ref{w-relation}), if we now insist, as we must, on retaining the causal
relationships among possible occasions in ${\cal O}$ classified above, then, the usual
positive definite product metric of the space ${\cal O}$ would have to be replaced by a
pseudo-Euclidean {\it indefinite} metric defined by
\begin{equation}
d{\css t2A}:=d{\css t2N} - {\css t2P}d{\bf{y}}^2,\label{a-metric}
\end{equation}
or, in Planck units, by
\begin{equation}
d{\css t2H}:=d{\css t2N} - d{\bf{y}}^2=:-\zeta_{\alpha\beta}\,dy^{\alpha}dy^{\beta}
\,,\label{a-Planck}
\end{equation}
together with the causality condition
\begin{equation}
d{\css t2A}\geq 0\label{A-causality}
\end{equation}
analogous to (\ref{causality}), where the line element ${d{\bf{y}}^2}$ is defined in
(\ref{y-metric}), and, for a later use, we have renamed ${d{\cs tA}}$ in (\ref{a-Planck})
by ${d{\cs tH}}$, with subscript ${H}$ standing for `Hamiltonian' (or `internal').
Thus, if the causal upper bound (\ref{t-bound}) is to be invariantly respected, then, even
in the rest frame, the {\it actual} proper duration registered by the clock would be
${d{\cs tA}}$, and not the usually supposed interval ${d{\cs tN}}$. Therefore, it is the time
${\cs tA}$, and not the Newtonian time ${\cs tN}$, that should appear in the Hamiltonian
flow equations, such as (\ref{H-eqn}), governing the evolution of states. In the
resulting mechanics, different canonical coordinates that are evolving with nonzero relative
rates would differ in general over which occasions are simultaneous with a given occasion. 
As unorthodox as this resulting picture may appear, it is an inevitable consequence of the
upper bound (\ref{t-bound}). 

Let us now return to the simple thought experiment considered in the previous section and ask:
In the clock's rest frame, what will be the time interval registered by it as it evolves from
occasion ${o_1:=(t_1,\,s_1)}$ to occasion ${o_2:=(t_2,\,s_2)}$ within the space
${({\cal O},\,\zeta)}$?
According to the line element (\ref{a-metric}), the answer clearly is
\begin{equation}
\Delta {\cs tA}=\int_{o_1}^{o_2}\frac{1}{\gamma(\omega)}\;d{\cs tN}
=\frac{\left|t_2-t_1\right|}{\gamma(\omega)}\,,
\end{equation}
were
\begin{equation}
1<\gamma(\omega):=\left(1-{\css t2P}\,\omega^2\right)^{-\frac{1}{2}}.
\end{equation}
Thus, if the state of the clock is at all evolving, we will have the phenomenon of `time
dilation' even in the rest frame. In particular, for a clock evolving at the
rate ${\css t{-1}P}$, the time stands still!
Similarly, we may now speak of a phenomenon of `{\it state}
contraction' in analogy with the familiar phenomenon of `length contraction': 
\begin{equation}
\Delta {\rm y}'=\omega\,\Delta{\cs tA}=\frac{\;\omega\,\Delta {\cs tN}}
{\gamma(\omega)}=\frac{\Delta {\rm y}}{\gamma(\omega)}\,.
\end{equation}
It is worth emphasizing here that, as in ordinary special relativity, nothing is actually
`dilating' or `contracting'. All that is being exhibited by these phenomena is that the two
sets of mutually evolving
canonical coordinates happen to differ over which occasions are simultaneous.

In addition to the light cones in spacetime bounded by the maximum possible speed ${c}$, we may
now also speak of {\it Planck cones} in the ${(1+2n)}$-dimensional space ${({\cal O},\,\zeta)}$
bounded by the maximum rate of evolution ${\css t{-1}P}$. Analogous to a light cone delimiting
the allowed causal relations between a given event and other events in spacetime
(cf. \cite{Wald-1984}), a Planck cone would delimit the causal relations between a given
occasion ${o_i}$ and other occasions within the space ${({\cal O},\,\zeta)}$. Accordingly, the
occasions that lie on the boundary of the `future Planck cone' of ${o_i}$ defined by
${\css t{-1}P}$ can be reached by a physical state at ${o_i}$ if and only if it is evolving at
the maximum possible rate ${\css t{-1}P}$. These occasions, and similarly defined occasions
lying on the `past Planck cone' of ${o_i}$, form a ${2n}$-dimensional set within
${({\cal O},\,\zeta)}$, and may be said to be {\it Planck-like} related to ${o_i}$. Occasions
that are on neither the past nor the future Planck cone of ${o_i}$ but lie interior to the
Planck cones are also causally accessible by a physical state at
${o_i}$, and may be said to be {\it time-like} related to ${o_i}$. The remaining occasions lying
exterior to the Planck cones comprise a causally forbidden ${\!(1\!+2n)}$-dimensional set within
${({\cal O},\,\zeta)}$, and may be said to be {\it state-like} related to ${o_i}$.

\section{proper time in a `moving frame'}

So far we have used a specific Lorentz frame, namely the rest frame of the clock, to obtain
expression (\ref{a-metric}) for the actual proper duration. For a frame with respect to which
the same clock is uniformly moving, the expression for actual proper duration can be now
obtained at once by using the Minkowski line element (\ref{e-metric}), yielding
\begin{equation}
d{\css t2A}=d{\css t2N}-c^{-2}d{\bf{x}}^2-{\css t2P}d{\bf{y}}^2.
\label{time1-metric}
\end{equation}
If we now eliminate the explicit appearance of the speed of light ${c}$ by defining it as a
Planck scale ratio ${{\cs lP}/{\cs tP}}$, then this actual proper duration can be rewritten as
\begin{equation}
d{\css t2A}=d{\css t2N}-{\css t2P}d{\bf{z}}^2,
\label{time2-metric}
\end{equation}
with the dimensionless ${3+2n}$ vector ${\bf z}$ defined by
\begin{equation}
{\bf z}:={\css l{-1}P}{\bf{x}}+{\bf{y}}\,.\label{z}
\end{equation}
The answer to the central question raised in Sec. III is now clear:
The time interval actually registered by the clock as it moves and evolves from the
event-state ${({\rm e}_1,\,{\rm s}_1)}$ to the event-state ${({\rm e}_2,\,{\rm s}_2)}$
in the combined space ${\cal E}$ is given, not by (\ref{not-this}) as customarily assumed,
but by
\begin{equation}
\Delta {\cs tA}=\int_{({\rm e}_1,\,{\rm s}_1)}^{({\rm e}_2,\,{\rm s}_2)}
\frac{1}{\gamma(\theta)}\;d{\cs tN}
=\frac{\left|t_2-t_1\right|}{\gamma(\theta)}\,,\label{z-dilation}
\end{equation}
were, in accordance with the line element (\ref{time2-metric}), 
\begin{equation}
1<\gamma(\theta):=\left(1-{\css t2P}\,\theta^2\right)^{-\frac{1}{2}},\label{z-gamma}
\end{equation}
with
\begin{equation}
\theta := \left|\frac{d{\bf z}({\cs tN})}{d{\cs tN}}\right| =
\left|{\css l{-1}P}{\bf v}+{\boldsymbol{\omega}}\right|.\label{Omega}
\end{equation}
This answer for the actual proper duration registered by the clock between the event-states
${({\rm e}_1,\,{\rm s}_1)}$ and ${({\rm e}_2,\,{\rm s}_2)}$ comprises {\it the} central,
empirically verifiable, claim of the theory proposed here. The fundamental reason behind
such an overall `time dilation' effect encoded in this answer
is the fact that, according to the present theory,
different inertial observers disagree in general, not just over which {\it events} occur
simultaneously with a given {\it event}, but over which {\it event-states} occur
simultaneously with a given {\it event-state}. Note, however, that away from the Planck scale,
and/or for slow evolutions of states, the ameliorated gamma factor (\ref{z-gamma}) duly
reduces to the usual gamma factor (\ref{x-gamma}) of special relativity, which of course
presumes absolute simultaneity of physical states.

We are now in a position to isolate the two basic postulates on which the theory developed
here can be erected in the manner analogous to the usual special relativity. In fact, the
first of the two postulates can be taken to be Einstein's very own first postulate, as stated
above, except that we must now revise what we mean by an inertial coordinate system. In normal
practice, an inertial coordinate system is taken to be a system of four dimensions,
one temporal and three spatial, moving uniformly in spacetime. In the present theory it
is understood to be a system of ${4+2n}$ dimensions, `moving' uniformly in the combined space
${\cal E}$, with four being the external spacetime dimensions, and ${2n}$ being the internal
phase space dimensions of a system under scrutiny. Again, the internal dimensions of the object
system can be either finite or infinite in number. The principle of relativity may now be
restated as follows: (i${'}$) {\it The laws governing the states of physical systems are
insensitive to `the state of motion', in the space ${\cal E}$, of the ${(4+2n)}$-dimensional
reference coordinate system, as long as it remains `inertial'}.

The second postulate on which the present theory is based is a generalization of the above
stated second postulate of Einstein. According to it: (ii${'}$) {\it No time rate of change of
a dimensionless physical quantity expressed in Planck units can exceed the inverse of the Planck
time}. In particular, for the dimensionless quantity {\bf z} defined above, this postulate may
be succinctly restated as
\begin{equation}
\theta \leq {\css t{-1}P},
\label{dimless}
\end{equation}
with understanding that this is a statement true for all local inertial
observers regardless of their state of motion. That is to say, in order for this causal upper
bound to be invariantly meaningful, the time rate of change
${\theta^I}$ of any such quantity ${{\rm z}^I}$ must satisfy the composition law
\begin{equation}
\theta'^I=\frac{\theta_{}^I+\,\theta^I_r}{1+{\css t2P}\;\theta_{}^I\,\theta^I_r}\,
\label{Omega-relation}
\end{equation}
analogous to (\ref{w-relation}), where ${I=1,2,\dots}$, or ${3+2n}$. 

Note that away from the Planck scale and/or when ${\boldsymbol{\omega}}$ is negligible,
${{\boldsymbol{\theta}}\approx{\css l{-1}P}{\bf v}}$, and (\ref{dimless}) duly reproduces
(\ref{c-bound}). Moreover, for external directions, ${I=1,2}$, or ${3}$,
(\ref{Omega-relation}) is identical to the composition law (\ref{v-relation}) for
velocities, and for internal directions, ${I=4,5,\dots}$, or ${3+2n}$, it
is identical to the similar composition law (\ref{w-relation}) for evolution rates. This means,
in particular, that, despite the generalization, the constant `${c}$' still remains
an observer-independent upper bound on admissible speeds. This invariance of `${c}$' will be
reaffirmed later (cf. eq. (\ref{reaffirmed})) as a derivative notion.

With the composition law (\ref{Omega-relation}), it is now easy to see that, just as in
Einstein's special relativity, the above two revised postulates inevitably lead to the actual
proper duration (\ref{time2-metric}), along
with the causality condition (\ref{A-causality}). For the sake of conceptual clarity,
so far in these
expressions we have used the notations ${\cs tE}$ and ${\cs tN}$ for the proper and coordinate
times, respectively. Employing more
familiar notations for these notions of time and using Planck
units, we now rewrite expression (\ref{time2-metric}) as
\begin{equation}
d\tau^2=\,dt^2-\,d{\bf z}^2=:\,-\,\xi_{AB}\,dz^A dz^B,\label{quad}
\end{equation}
where the index ${A=0,\dots,3+2n}$ runs along the ${4+2n}$ dimensions of the
pseudo-Euclidean manifold ${({\cal E},\,\xi)}$. This, then, according to the present theory,
is the true quadratic invariant of the inertial structure of the world.

\section{Refined local inertial physics}

Having arrived at the above refined chart of the inertial structure, we shall now see how it is
further justified by its amicable theoretical consequences. In particular, we shall see how, in
the theory based on the refined structure, energies and momenta turn out to be invariantly
bounded from above, and lengths and durations similarly bounded from below, by their respective
Planck scale values.

\subsection{Permitted coordinate transformations}

The coordinate transformations in the combined space ${({\cal E},\,\xi)}$ analogous to the
Lorentz transformations in the Minkowski spacetime that preserve the
above fundamental quadratic invariant ((\ref{quad}) or (\ref{time1-metric})) can be written as
\begin{equation}
z^A=\Lambda^A_{\,\;\;B}\,z'^B +\,b^A\,,\label{transla}
\end{equation}
where ${b^A}$ and ${\Lambda^A_{\,\;\;B}}$ are constants, constrained by
\begin{equation}
\Lambda^A_{\,\;\;C}\,\Lambda^B_{\,\;\;D}\;\xi_{AB}=\xi_{CD}\,.\label{consla}
\end{equation}
At least for simple, finite dimensional, phase spaces, the coefficients ${\Lambda^A_{\,\;\;B}}$
can be easily determined. Consider, for example, a massive relativistic particle at rest (and
hence also not evolving) with respect to a primed coordinate system in the external spacetime,
which is moving with a uniform velocity ${\bf v}$ with respect to another unprimed coordinate
system. Since, as it moves, the state of the particle will also be evolving in its six
dimensional phase space (and since the evolution of a mechanical system is simply the
continuous unfolding of canonical transformations \cite{Arnold-1989}), we can view its motion
and evolution together with respect to a ${(4+6)}$-dimensional unprimed coordinate system in the
space ${({\cal E},\,\xi)}$, and thus view it to be moving and evolving, say, at an arbitrary
combined rate ${{\boldsymbol{\theta}}}$ (cf. (\ref{Omega})). 
Now, from (\ref{transla}) we have the differential relations 
\begin{equation}
dz^A=\Lambda^A_{\,\;\;B}\,dz'^B\,.\label{difla}
\end{equation}
Since ${d{\bf z}'}$ vanishes in the present case, these reduce to
\begin{align}
d{\rm z}^I \!& =\Lambda^I_{\;\;0}\,dt'\;\;\;\;\;\;\;(I=1,2,\dots,9=3+6),\label{reduz}\\
{\rm and}\;\;\;\;\;\;\;dt & =\Lambda^0_{\;\;0}\,dt'\,.\label{redut}
\end{align}
Now, dividing (\ref{reduz}) by (\ref{redut}) we have one relation
between the coefficients ${\Lambda^I_{\;\;0}}$ and ${\Lambda^0_{\;\;0}\,}$,
\begin{equation}
\Lambda^I_{\;\;0}=\theta^I\Lambda^0_{\;\;0}\,,
\end{equation}
and setting ${C=D=0}$ in (\ref{consla}) we have another,
\begin{equation}
-1=\Lambda^A_{\;\;0}\Lambda^B_{\;\;0}\,\xi_{AB}= - (\Lambda^0_{\;\;0})^2 +
\sum_{I=1}^{3+6}(\Lambda^I_{\;\;0})^2\,.
\end{equation}
The solution of these two simultaneous equations is
\begin{equation}
\Lambda^0_{\;\;0}=\gamma(\theta)\;\;\;\;\;{\rm and}\;\;\;\;\;
\Lambda^I_{\;\;0}=\gamma(\theta)\,\theta^I\,,
\end{equation}
with ${\gamma(\theta)}$ given by (\ref{z-gamma}). As in the familiar case of Lorentz
transformations, hear also the other ${\Lambda^A_{\,\;\;B}}$ can be determined
uniquely only up to arbitrary rotations. One convenient choice that would satisfy
the constraint (\ref{consla}) is
\begin{equation}
\Lambda^0_{\;\,J}=\gamma(\theta)\,\theta^{}_J \;\;\;{\rm and}\;\;\;
\Lambda^I_{\;\,J}=\delta^I_{\;J}+\frac{\theta^I_{}\theta^{}_J}{{\boldsymbol{\theta}}^2}
\left[\gamma(\theta)-1\right].\label{other}
\end{equation}

\subsection{Bounds on lengths and durations}

Let us now restrict to the following two very special cases of only external spatio-temporal
transformations:
\begin{align}
d{\rm x}^k\! & =\gamma(v,\,\omega)\,d{\rm x}'^k
\;\;\;\;\;(k=1,2,\;{\rm or}\;\,3),\label{reduzsol}\\
{\rm and}\;\;\;\;\;\;\;dt & =\gamma(v,\,\omega)\,dt',\label{redutsol}
\end{align}
where, reactivating the units for clarity, we have written
\begin{equation}
\gamma(\theta)\equiv\gamma(v,\,\omega)
=\left(1-c^{-2}\,v^2 - {\css t2P}\,\omega^2\right)^{-\frac{1}{2}}.\label{gamma-oz}
\end{equation}
Using (\ref{Hamilton-Jacobi}), this gamma factor can also be written as
\begin{equation}
\gamma(v,\,\omega)\equiv\gamma(v,\,dE)=
\left[1-\frac{v^2}{c^2} - \frac{\;\left(dE\right)^2}{\css E2P}
\right]^{-\frac{1}{2}}\!.\label{gamma-E-inf}
\end{equation}
If we now assume that the difference of particle energy between
the two coordinate systems is small but finite, that is, if we assume that
\begin{equation}
\frac{dE}{\;{\cs EP}}\sim\frac{\Delta E}{\;{\cs EP}}=
\frac{E'-E}{\;{\cs EP}}\,,\label{differ-rela}
\end{equation}
then the gamma factor (\ref{gamma-E-inf}) takes the useful form
\begin{equation}
\gamma(v,\,\omega)\sim\gamma(v,\,\Delta E)=
\left[1-\frac{v^2}{c^2} - \frac{\;\left(E'-E\right)^2}{\css E2P}
\right]^{-\frac{1}{2}}\!.\label{gamma-E-fin}
\end{equation}

Now (\ref{reduzsol}) and (\ref{redutsol}) are clearly the infinitesimal counterparts of
the refined `length contraction' and `time dilation' relations. For a small but finite distance
in the direction of the relative
velocity ${\bf v}$, the first of these two relations can be rewritten as
\begin{equation}
\Delta {\rm x}'=\Delta {\rm x}\,\sqrt{1-\frac{v^2}{c^2} -
\,{\css l2P}\left(\frac{{\Delta {\rm x}}-{\Delta {\rm x}'}}{{\Delta {\rm x}'}{\Delta {\rm x}}}
\right)^2}\,.\label{dist}
\end{equation}
Here we have dropped the index on ${\Delta{\rm x}}$ and used
\begin{equation}
\frac{E'-E}{\;{\cs EP}}=\frac{\,{\cs lP}}{\Delta {\rm x}'}-
\frac{\,{\cs lP}}{\Delta {\rm x}}\;,\label{hj-rela}
\end{equation}
which follows from an inversion of units analogous to the one routinely used in high energy
physics. The `length contraction' expression (\ref{dist}) can be exactly solved for the
`contracted' length ${\Delta {\rm x}'}$ in terms of the `uncontracted' length
${\Delta {\rm x}}$. However, the solutions of (\ref{dist}) provided by the computer software
Maple are far too long and complicated to be reproduced here. Fortunately, exact solutions of
(\ref{dist}) are not necessary for our purposes of demonstrating the existence of a lower bound
on lengths.

We begin with the assumption that the `contracted' length ${\Delta {\rm x}'}$ is negligibly
smaller than the `uncontracted' length ${\Delta {\rm x}}$. The question then is, how small the
`contracted' length can get? In other words, is there an absolute lower bound on lengths?
The answer, according to the present theory, is, yes, there is a lower bound, and it
is no other than ${\cs lP}$. To see this, note that under the assumption
${\Delta {\rm x}'\ll\Delta {\rm x}}$, the relation (\ref{dist}) reduces to
\begin{equation}
\Delta {\rm x}'=\Delta {\rm x}\,\sqrt{1-\frac{v^2}{c^2} -
\,\frac{{\css l2P}}{\left(\Delta {\rm x}'\right)^2}}\,.\label{dist-2}
\end{equation}
Squaring both sides of this equation yields
\begin{equation}
(\Delta {\rm x}')^4 - \left(1-\frac{v^2}{c^2}\right)
(\Delta {\rm x})^2(\Delta {\rm x}')^2+ (\Delta {\rm x})^2\,{\css l2P}=0\,.\label{sqr}
\end{equation}
The only positive real root of this quartic equation for ${\Delta {\rm x}'}$ that gives the
correct physical limit 
\begin{equation}
\lim_{\Delta {\rm x}\gg\,{\cs lP}}
\Delta {\rm x}'=\Delta {\rm x}\,\sqrt{1-\frac{v^2}{c^2}}=\frac{\Delta {\rm x}}{\gamma(v)}
\label{lim-l}
\end{equation}
is the following expression for `length contraction',
\begin{equation}
\Delta {\rm x}'=\Delta {\rm x}\,\sqrt{\frac{1}{2}\left(1-\frac{v^2}{c^2}\right) +
{\sqrt{\frac{1}{4}\left(1-\frac{v^2}{c^2}\right)^{\!2} -
\frac{{\css l2P}}{(\Delta {\rm x})^2}}}}\;,\label{l}
\end{equation}
provided the reality condition
\begin{equation}
\frac{1}{4}\left(1-\frac{v^2}{c^2}\right)^{\!2}\;\geq\;
\frac{{\css l2P}}{(\Delta {\rm x})^2}\label{inequal}
\end{equation}
is satisfied. Substituting this last inequality back into the solution (\ref{l}) then gives
\begin{equation}
\Delta {\rm x}'\;\geq\;\sqrt{{\cs lP}\Delta {\rm x}}\;.\label{bound-pre}
\end{equation}
Thus, as long as ${\Delta {\rm x}}$ is chosen to be greater than ${\cs lP}$, the `contracted'
length ${\Delta {\rm x}'}$
also remains greater than ${\cs lP}$. That is to say, along with the upper
bound implied by the condition ${\gamma(v,\,\omega)>1}$, the `contracted' length remains
invariantly bounded from below as well as from above:
\begin{equation}
\Delta {\rm x}\;>\;\Delta {\rm x}'\;>\;{\cs lP}\,.\label{bound-l}
\end{equation}

Starting again from the infinitesimal expression (\ref{redutsol})---now for
`time dilation'---and using almost identical line of arguments as the case above (along with
the assumption ${\Delta\tau\ll\Delta t}$ for the `dilated' time), we arrive,
in analogous manner, at a refined expression for `time dilation',
\begin{equation}
\Delta\tau=\Delta t\,\sqrt{\frac{1}{2}\left(1-\frac{v^2}{c^2}\right) +
{\sqrt{\frac{1}{4}\left(1-\frac{v^2}{c^2}\right)^{\!2} -
\frac{{\css t2P}}{(\Delta t)^2}}}}\;,\label{t}
\end{equation}
together with the corresponding invariant bounds on the `dilated' or proper time ${\Delta\tau}$:
\begin{equation}
\Delta t\;>\;\Delta\tau\;>\;{\cs tP}\,.\label{bound-t}
\end{equation}

\subsection{Reaffirming the bound on velocities}

So far we have not assumed or proved explicitly that the constant `${c}$' is an upper bound on
possible speeds. As emphasised before (cf. comments after (\ref{Omega-relation})), in the
present theory the upper bound `${c}$' and its observer-independence turn out to be derivative
notions. To see this explicitly, consider the ratio of `contracted' length (\ref{l}) and
`dilated' duration (\ref{t}); that is, consider
\begin{equation}
u'=\,u\,\sqrt{\frac{\frac{1}{2}\left(1-\frac{v^2}{c^2}\right) +
{\sqrt{\frac{1}{4}\left(1-\frac{v^2}{c^2}\right)^{\!2} -
\frac{{\css l2P}}{(\Delta {\rm x})^2}}}}
{\frac{1}{2}\left(1-\frac{v^2}{c^2}\right) +
{\sqrt{\frac{1}{4}\left(1-\frac{v^2}{c^2}\right)^{\!2} -
\frac{{\css t2P}}{(\Delta t)^2}}}}}\;,\label{ratio}
\end{equation}
where we have defined `velocities' ${u}$ and ${u'}$ as
\begin{equation}
u:=\frac{\Delta {\rm x}}{\Delta t}\;\;\;\;\;\;{\rm and}\;\;\;\;\;\;
u':=\frac{\Delta {\rm x}'}{\Delta\tau}\,.\label{u-defined}
\end{equation}
It is clear from these definitions that, for an arbitrary ${\Delta {\rm x}'}$ subject to the
bounds (\ref{bound-l}), ${u'}$ would be the largest possible velocity when ${\Delta\tau}$
happens to be the smallest possible duration, which, in turn, for arbitrary but finite ${v}$ and
${\Delta t}$ (according to (\ref{t})), is given by the condition
\begin{equation}
\frac{1}{4}\left(1-\frac{v^2}{c^2}\right)^{\!2}=\frac{{\css t2P}}{(\Delta t)^2}.\label{small-t}
\end{equation}
Consequently, substituting this condition
back into the ratio (\ref{ratio}) gives the following upper bound on ${u'}$:
\begin{equation}
u'\,\leq\,u\,\sqrt{1+\sqrt{1-c^2\,u^{-2}}}\,.\label{reaffirmed}
\end{equation}
Thus, as long as ${u}$ does not exceed ${c}$ (and the resulting upper bound remains real),
${u'}$ also remains within ${c}$. In other words, in the present theory ${c}$ retains its
usual status of the observer-independent upper bound on causally admissible speeds, albeit in a
rather derivative manner.

\subsection{Elements of particle mechanics}

Let us now take the mass of the particle considered above to be ${m}$. Then, just as
in the standard special relativity \cite{Taylor-1992}, the fundamental quadratic
invariant (\ref{quad}) leads to the `momentum space' quadratic invariant
\begin{equation}
\xi_{AB}{\cal P}^A{\cal P}^B = -\,m^2\,c^2\,,\label{2n-mom-inv}
\end{equation}
provided we define the ${4+2n}$ momentum ${\boldsymbol{\cal P}}$ as
\begin{equation}
m\,\frac{\;dz^A}{d\tau\,}=:{\cal P}^A := \left(E/c\,,\,P^I\right),\label{4+2n-mom}
\end{equation}
where the energy of the particle is now defined as
\begin{equation}
E := \gamma(v,\,dE)\;m\,c^2\,,\label{energy}
\end{equation}
with ${\gamma(v,\,dE)}$ given by (\ref{gamma-E-inf}),
and the ${3+2n}$ momentum ${P^I}$ (${I=1,2,\dots,3+2n}$) of the particle is defined as
\begin{equation}
{\bf P}:={\bf P}_{ext}+{\bf P}_{int}\,,\label{int-ext-mom}
\end{equation}
with
\begin{equation}
{\bf P}_{ext}\equiv{\bf p} := \gamma(v,\,dE)\;m\,{\bf v}\label{ext-mom}
\end{equation}
and
\begin{equation}
{\bf P}_{int}:= \gamma(v,\,dE)\;m\,{\cs lP}{\boldsymbol{\omega}}\,.\label{int-mom}
\end{equation}
Clearly, apart from the modified gamma factor, the expression (\ref{energy}) of energy is no
different from the familiar one. The expression (\ref{int-ext-mom}) for momentum, on the other
hand, has two parts: an `external' part ${{\bf P}_{ext}}$ and an `internal' part
${{\bf P}_{int}}$. Again, apart from the modified gamma factor, the external momentum
${{\bf P}_{ext}}$ is no different from the familiar one, and hence in what follows it will be
denoted simply by ${\bf p}$. On the other hand, the notion of the internal ${2n}$ momentum
${{\bf P}_{int}}$ is novel, and its physical meaning is as follows: Recall that
${\boldsymbol{\omega}}$ in (\ref{int-mom}) is simply the Hamiltonian vector field, which Arnold
\cite{Arnold-1989} occasionally refers to as `the phase velocity vector field'. In the similar
manner, ${{\bf P}_{int}}$ may be referred to as {\it the phase momentum vector field} on
${\cal N}$. Its usefulness will become clear soon.

Substituting now the explicit expression for ${\boldsymbol{\cal P}}$ from (\ref{4+2n-mom})
into the quadratic invariant (\ref{2n-mom-inv}), we obtain the following modified expression
for the dispersion relation:
\begin{equation}
|{\bf p}|^2\,c^{\,2}\,+\;m^2\,c^4\,=\,E^2\,-\,|{\bf P}_{int}|^2\,c^2\,.\label{int-disp}
\end{equation}
Using the definitions (\ref{int-mom}) for ${{\bf P}_{int}}$ and (\ref{energy}) for energy,
along with the equations (\ref{one-form}) and (\ref{Hamilton-Jacobi}) for dynamics, the
1-form associated with the vector field ${{\bf P}_{int}}$ (which we denote also by
${{\bf P}_{int}}$ for convenience) can be written as 
\begin{equation}
{\bf P}_{int}=\frac{E\,{\bf d}E}{c\,{\cs EP}}\,.
\label{trans-int}
\end{equation}
Substituting this into (\ref{int-disp}), we finally arrive at the refined dispersion
relation between energies and momenta:
\begin{equation}
p^2\,c^2\,+\;m^2\,c^4\,=\,E^2\left[\,1\,-\,\frac{\;\left(dE\right)^2}{\css E2P}\,\right].
\label{exact-disp}
\end{equation}
It is worth noting that this is an {\it exact} relation between energies and momenta. We shall
return to it in Sec. VIII to discuss how it can be experimentally verified. As a
consistency check, note also that in the rest frame of the massive particle both ${\bf p}$ and
${{\bf d}E}$ vanish identically, and (\ref{exact-disp}) reproduces the famous
mass-energy equivalence:
\begin{equation}
E=m\,c^2.\label{mass-energy}
\end{equation}
If we now define {\it refined energy} ${\cs ER}$ as
\begin{equation}
{\cs ER}\,:=\,E\,\sqrt{1\,-\,\frac{\;\,(dE)^2}{\css E2P}\,},
\label{refi-energy}
\end{equation}
then (\ref{exact-disp}) takes the following perspicuous form:
\begin{equation}
{\css E2R}\,=\,p^2\,c^2\,+\;m^2\,c^4.
\label{perspi-disp}
\end{equation}
Since ${{\cs ER}\rightarrow E\rightarrow m\,c^2}$ as ${p\rightarrow 0}$, in the present theory
this relation plays the role similar to that played by the usual dispersion relation in special
relativity.
 
\subsection{Covariant conservation laws}

Consider an isolated system of mass ${m_{sys}}$ composed of a number of constituents undergoing
an internal reaction. In view of the quadratic invariant (\ref{2n-mom-inv}), and considering
what we have learned from special relativity \cite{Taylor-1992},
it is only natural to assume that the ${4+2n}$ momentum
${{\boldsymbol{\cal P}}_{\!sys}}$ of the system would be conserved in such a reaction,
\begin{equation}
\Delta {\boldsymbol{\cal P}}_{\!sys}=0\,,\label{totmomcons}
\end{equation}
where ${\Delta}$ denotes the difference between the initial and final states of the reaction,
and ${{\boldsymbol{\cal P}}_{\!sys}}$ is defined by
\begin{equation}
m_{sys}\,\frac{\;dz^A}{d\tau\,}=:{\cal P}_{\!sys}^A := 
\left(E_{sys}/c\,,\;p_{sys}^k\,,\;P^{\mu}_{sys}\right),\label{4+2n-sys}
\end{equation}
with ${k=1,2,3}$ being the external three dimensions and ${\mu=4,5,\dots,3+2n}$ being the phase
space dimensions of the system as a whole. It is clear from this definition that, since
${dz^A}$ is a ${(4+2n)}$-vector whereas ${m_{sys}}$ and ${d\tau}$ are invariants,
${{\cal P}_{\!sys}^A}$ is also a ${(4+2n)}$-vector, and hence transforms under (\ref{difla}) as
\begin{equation}
{\cal P}_{\!sys}'^A=\Lambda^A_{\,\;\;B}\;{\cal P}_{\!sys}^B\,.\label{transla-mom}
\end{equation}
Moreover, since ${\Lambda}$ does not depend on anything but the permitted coordinate
transformation being performed in the space ${({\cal E},\,\xi)}$, the difference on the left
hand side of (\ref{totmomcons}) is also a ${\!(4+2n)}$-vector, and therefore transforms as
\begin{equation}
\Delta {\cal P}_{\!sys}'^A=\Lambda^A_{\,\;\;B}\;\Delta
{\cal P}_{\!sys}^B\,.\label{transla-differ}
\end{equation}
Thus, if the conservation law (\ref{totmomcons}) holds for one set of ${(4+2n)}$-dimensional
coordinates in ${({\cal E},\,\xi)}$, then, according to (\ref{transla-differ}), it does so
for {\it all} ${(4+2n)}$-dimensional
coordinates related by the transformations (\ref{transla}). Care must be taken, however, when
implementing such a deceptively simple conservation law in a real physical reaction.

To appreciate one of the main subtleties,
let us unpack (\ref{totmomcons}) into its external, internal, and constituent parts:
\begin{equation}
0=\Delta {\boldsymbol{\cal P}}_{\!sys}=\left(\Delta E_{sys}/c\,,\;
\Delta {\bf p}_{sys}\,,\;\Delta {\bf P}^{sys}_{int}\right),\label{extmomcons}
\end{equation}
with
\begin{equation}
\Delta E_{sys}:=\sum_f E_f -\sum_i E_i\label{sumenegy}
\end{equation}
and
\begin{equation}
\Delta {\bf p}_{sys}:=\sum_f {\bf p}_f -\sum_i {\bf p}_i\,,\label{summom}
\end{equation}
where the indices ${f}$ and ${i}$ stand for the final and initial number of constituents of the
system. Thus, as in the special theory of relativity (and unlike in the two-scales theories
\cite{Judes}), in the present theory energies and momenta remain
{\it additive}. What is crucial to note, however, is that ${{\bf P}^{sys}_{int}\,}$,
being the phase momentum vector field of the system as a whole, {\it does not}, in general,
warrant decomposition analogous to (\ref{summom}) into constituent parts.

Another aspect the conservation law (\ref{totmomcons}) worth noting is that all of its
${4+2n}$ components stand or fall together: any one of them entails all the others. Since
${\Delta {\boldsymbol{\cal P}}_{\!sys}}$ is a ${(4+2n)}$-vector, the relativity principle
in the form of the coordinate transformations (\ref{transla-differ}) dictates that, if any
one of the components of the ${(4+2n)}$-momentum ${{\boldsymbol{\cal P}}_{\!sys}}$ is
invariantly conserved
in a reaction, then the entire vector, with all of its components, must also be so conserved.
In particular, if the energy ${E_{sys}}$ is conserved in a reaction, then so is the momentum
${{\bf p}_{sys}}$, and vise versa.

Sometimes it is asserted \cite{Amelino-Camelia-2003} that the desire to maintain unmodified
energy-momentum conservation laws accompanying a modified dispersion relation such as
(\ref{exact-disp}) may come at a price of introducing a preferred class of inertial observers
in a theory; and, conversely, avoiding such a preferred class of observers would necessarily
lead to a modification of conservation laws. However, it is clear form the above discussion
that, in the present theory, not only are there no preferred class of observers, but also the
conservation laws of special relativity---at least in the external sector---remain
essentially unchanged.

\subsection{Bounds on energies and momenta}

Not surprisingly,
analogous to the lower bounds on lengths and durations, in the present theory energies and
momenta are also invariantly bounded from above. In the case of energy, this can
be seen from the definition of energy (\ref{energy}) itself. Using (\ref{gamma-E-fin}), and an
assumption similar to the one we used for the `contracted' length (again employed for the same
practical reasons), namely ${E'\gg E}$, the definition (\ref{energy}) of energy can be
expressed as
\begin{equation}
E'=\frac{E}{\sqrt{1-\frac{v^2}{c^2} - \frac{\;\left(E'\right)^2}{\css E2P}}}\,.\label{redu-engy}
\end{equation}
Squaring both sides of this equation leads to the following quartic equation for energy:
\begin{equation}
E'^{\,4}-\left(1-\frac{v^2}{c^2}\right)E'^{\,2}{\css E2P}+ E^2{\css E2P}=0\,.\label{sqr-E}
\end{equation}
The only positive real root of this equation for ${E'}$ that gives the correct physical limit 
\begin{equation}
\lim_{E\ll\,{\cs EP}}
E'=E\left(1-c^{-2}v^2\right)^{\!-\frac{1}{2}}={\gamma(v)}E\label{lim-e}
\end{equation}
is the following refined expression for energy:
\begin{equation}
E'=E\,\sqrt{\left(1-\frac{v^2}{c^2}\right)\frac{{\css E2P}}{2E^2}-{
\sqrt{\left(1-\frac{v^2}{c^2}\right)^{\!2}
\frac{{\css E4P}}{4E^4}-\frac{{\css E2P}}{E^2}\,}}}\,,\label{e}
\end{equation}
provided the reality condition
\begin{equation}
\left(1-\frac{v^2}{c^2}\right)^{\!2}
\frac{{\css E4P}}{4E^4}\;\geq\;\frac{{\css E2P}}{E^2}\label{E-inequal}
\end{equation}
is satisfied. Substituting this last inequality back into the solution (\ref{e}) then gives
\begin{equation}
E'\;\leq\;\sqrt{{\cs EP}E}\;.\label{E-bound-pre}
\end{equation}
Thus, as long as ${E}$ does not exceed ${\cs EP}$, ${E'}$ also remains within ${\cs EP}$.
That is, along with the lower bound implied by the condition ${\gamma(v,\,\Delta E)>1}$, the
energy remains invariantly bounded from both below and above:
\begin{equation}
E\;<\;E'\;<\;{\cs EP}\,.\label{bound-e}
\end{equation}

Note that, unlike the evaluations of limit (\ref{lim-l}), the closed form
evaluation of the limit (\ref{lim-e}) is not straightforward and requires an iterative use
of the L'H\^opital's rule. In other words, the dependence of ${E'}$ on the vanishing ratio
${E/{\cs EP}}$ is rather subtle compared to, say, the dependence of ${\Delta {\rm x}'}$ on
the analogous ratio ${{\cs lP}/\Delta {\rm x}}$.

Starting again from the definition (\ref{ext-mom}) for momentum, and using almost identical line
of arguments as the case above, we can arrive at an analogously refined expression for momentum
as follows. As in special relativity, the ratio of (\ref{ext-mom}) and (\ref{energy}) gives the
identity
\begin{equation}
\frac{p\,c}{E}=\frac{v}{c}\,,\label{p-E-identity}
\end{equation}
which, upon using the Planck momentum ${{\cs kP}:={\cs EP}/c}$, can also be written as
\begin{equation}
\frac{E}{\cs EP}=\frac{p\,c}{{\cs kP}v}\,.\label{Ep-kp}
\end{equation}
Using (\ref{gamma-E-fin}) and this identity into (\ref{ext-mom}) (along with the
assumption ${p'\gg p}$) and then squaring both sides of (\ref{ext-mom}) leads to the following
quartic equation for momentum:
\begin{equation}
p'^{\,4} -\left(1-\frac{v^2}{c^2}\right)\frac{v^2}{c^2}\,{\css k2P}\,p'^{\,2} +
\frac{v^2}{c^2}\,{\css k2P}\,p^2\,=\,0\,.\label{sqr-p}
\end{equation}
The only positive real root of this equation
that gives the correct physical limit 
\begin{equation}
\lim_{pv\,\ll\,{\cs kP}c}\,
p'\,=\,p\left(1-c^{-2}v^2\right)^{\!-\frac{1}{2}}={\gamma(v)}\,p\label{lim-p}
\end{equation}
is the following refined expression for momentum:
\begin{equation}
p'=p\sqrt{\!\left(1-\frac{v^2}{c^2}\right)\!\frac{{\css k2P}}{2p^2}\frac{v^2}{c^2}-{
\sqrt{\!\left(1-\frac{v^2}{c^2}\right)^{\!2}\!\!\frac{{\css k4P}}{4p^4}\frac{v^4}{c^4}-
\frac{{\css k2P}}{p^2}\frac{v^2}{c^2}}}}\,,\label{p}
\end{equation}
provided the reality condition
\begin{equation}
\left(1-\frac{v^2}{c^2}\right)^{\!2}
\frac{{\css k4P}}{4p^4}\frac{v^4}{c^4}\;\geq\;\frac{{\css k2P}}{p^2}\frac{v^2}{c^2}
\label{p-inequal}
\end{equation}
is satisfied. Substituting this last inequality back into the solution (\ref{p}) then gives
\begin{equation}
p'\;\leq\;\sqrt{{\cs kP}\,p}\,\times\sqrt{\frac{v}{c}}\;.\label{k-p-bound}
\end{equation}
Thus, as long as ${v}$ does not exceed ${c}$ and ${p}$ does not exceed ${\cs kP}$, ${p'}$ also
remains within ${\cs kP}$.
That is, along with the lower bound implied by the condition ${\gamma(v,\,\Delta E)>1}$, the
momentum remains invariantly bounded from above as well as from below:
\begin{equation}
p\;<\;p'\;<\;{\cs kP}\,.\label{bound-p}
\end{equation}

In summary, we conclude from the bounds (\ref{bound-l}), (\ref{bound-t}), (\ref{bound-e}), and
(\ref{bound-p}) that, in the present theory, {\it all} physical quantities of interest are
{\it invariantly} bounded by their respective Planck scale counterparts.

\subsection{Massless particles and Doppler shifts}

In Einstein's special relativity, Doppler shifts provide one of the most transparent
demonstrations of how the principle of relativity holds in nature. Despite the highly nonlinear
character of the `external' relations hidden in the overall linear transformations
(\ref{transla}), the same remains true here, albeit with Planck scale enhanced Doppler shifts.
In order to appreciate this, let us first spell out 
the general properties of massless particles in the theory.

It is evident from the invariant (\ref{2n-mom-inv}) that energy and momentum of any such
massless particle must satisfy
\begin{equation}
E^2\,=\,|{\bf P}|^2\,c^2\,\equiv\,|{\bf p}|^2\,c^2\,+|{\bf P}_{int}|^2\,c^2
\,,\label{mass-disp}
\end{equation}
which---as we saw in the case of the general dispersion relation (\ref{exact-disp})---is
equivalent to the relation
\begin{equation}
p^2\,c^2\,=\,E^2\left[\,1\,-\,\frac{(dE)^2}{\css E2P}\,\right].
\label{massless-disp}
\end{equation}
Using the identity (\ref{p-E-identity}), it is then easy to see that
the massless particle must move with the speed
\begin{equation}
v\,=\,c\;\sqrt{1-\frac{(dE)^2}{\css E2P}}\,.\label{fun-rela}
\end{equation}
Thus, any massless particle such as a photon with non-zero energy always moves with a speed
somewhat less than ${c}$, and if the difference ${dE}$ in energy happens to approach the
Planck energy ${\cs EP}$, then it hardly moves.

Consider now a receiver receding from a photon source with a uniform velocity ${\bf v}$. As a
given photon of energy ${E}$ propagates, it would also evolve in its phase space ${\cal N}$
with a uniform rate, say ${\boldsymbol{\omega}}$, relative to the canonical coordinates chosen
in ${\cal N}$ as a part of the reference frame for the source (recall that, in the present
theory, a frame of reference is a ${(4+2n)}$-dimensional object). If we now take the combined
relative `velocity' between the source and the receiver to be ${\boldsymbol{\theta}}$ as
defined by (\ref{Omega}), then the energy of the photon observed by the receiver would be simply
\begin{equation}
E'=\gamma(\theta)\left[E-{\cs lP}{\boldsymbol{\theta}}\cdot{\bf P}\,\right],
\label{photon-energy}
\end{equation}
which follows from a transformation analogous to (\ref{transla}) applied to the photon's 4+2n
momentum ${\boldsymbol{\cal P}}$ defined by (\ref{4+2n-mom}). If we next assume that the angle
between the ${(4+2n)}$-dimensional vectors ${\boldsymbol{\theta}}$ and ${\bf P}$ is ${\Phi}$,
then, as a result of the relation (\ref{mass-disp}), this transformation simplifies to
\begin{equation}
E'=\gamma(\theta)E\left[1-{\cs tP}\theta\,\cos\Phi\,\right].\label{photon-reduced}
\end{equation}
Of course, an external experimenter cannot be expected to have a direct access to either
${\theta}$ or ${\Phi}$, but, as we shall see, the knowledge of either
${\theta}$ or ${\Phi}$ is not necessary.

Now, to see how the principle of relativity works in the present theory, let us ask what would
be the energy of the photon observed by the receiver if we view the source to be receding rather
than the receiver. In other words, we now analyze the problem using the ${(4+2n)}$-dimensional
reference frame of the receiver rather than that of the source. The answer, not surprisingly,
is given by the inverse transformation
\begin{equation}
E=\gamma(\theta)\left[E'+{\cs lP}{\boldsymbol{\theta}}\cdot{\bf P}'\,\right],
\label{source-energy}
\end{equation}
which, similarly to (\ref{photon-reduced}), reduces to
\begin{equation}
E=\gamma(\theta)E'\left[1+{\cs tP}\theta\,\cos\Phi'\,\right],\label{source-reduced}
\end{equation}
where ${\Phi'}$ is the angle between the ${(4+2n)}$-dimensional vectors ${\boldsymbol{\theta}}$
and ${{\bf P}'}$. If we now make the substitutions ${\theta'^I=-{\css t{-1}P}\cos\Phi'}$,
${\theta^I=-{\css t{-1}P}\cos\Phi}$, and ${\theta^I_r=\theta}$ into the composition law
(\ref{Omega-relation}), then we expectedly arrive at the aberration relation between the
angles ${\Phi'}$ and ${\Phi}$:
\begin{equation}
\cos\Phi'=\frac{\cos\Phi-{\cs tP}\theta}{1-{\cs tP}\theta\cos\Phi}\,.
\label{aberration}
\end{equation}
Upon using this relation, along with a use of (\ref{z-gamma}), it is then easy to see that
(\ref{photon-reduced}) and (\ref{source-reduced}) are, in fact, identical relations between the
observed and source energies of the photon. Thus, as in special relativity, the principle of
relativity here as well renders it impossible to distinguish between the motions of the
source or the receiver.

As promised, let us now obtain the relation between the overall angle ${\Phi}$ and the external
angle, say ${\phi}$, which is the angle between the relative velocity ${\bf v}$ and photon
momentum ${\bf p}$. This relation between ${\Phi}$ and ${\phi}$ can be obtained by expanding
the dot product ${{\cs tP}{\boldsymbol{\theta}}\cdot{\bf P}}$ as
\begin{equation}
{\cs tP}\theta\cos\Phi=\frac{1}{|{\bf P}|}\left(c^{-1}v\,p\,\cos\phi\,+\,{\cs tP}\omega\,
|{\bf P}_{int}|\right)\,,\label{angle-dot}
\end{equation}
where we have used the fact that---as is evident from the definition
(\ref{int-mom})---${\boldsymbol{\omega}}$
and ${{\bf P}_{int}}$ are parallel vectors.
Using (\ref{mass-disp}), (\ref{massless-disp}), (\ref{energy}), (\ref{int-mom}),
(\ref{trans-int}), (\ref{Hamilton-Jacobi}), and (\ref{differ-rela}),
this expansion can be reduced to
\begin{equation}
{\cs tP}\theta\cos\Phi=\frac{(E'-E)^2}{\css E2P}+\left(\frac{v}{c}\cos\phi\right)
\sqrt{1-\frac{(E'-E)^2}{\css E2P}}\,.\label{angle-rela}
\end{equation}
Substituting this relation between the angles
${\Phi}$ and ${\phi}$ back into (\ref{photon-reduced})
and using (\ref{gamma-E-fin}) then gives the desired ratio
\begin{equation}
\frac{E'}{E\;}=\frac{\,\varepsilon'\left[\,\varepsilon'-\frac{v}{c}\cos\phi\,\right]}
{\sqrt{(\varepsilon')^2-\frac{v^2}{c^2}\,}\;},\label{newpho}
\end{equation}
where we have defined
\begin{equation}
\varepsilon':=\sqrt{1-\frac{E^2}{\css E2P}\!
\left(1-\frac{E'}{E\;}\right)^{\!2}}\,,\label{(1')}
\end{equation}
which clearly becomes unity for ${E\ll{\cs EP}}$, thus reducing (\ref{newpho}) to the
familiar expression for Doppler shifts. In fact, it can be shown that the formula (\ref{newpho})
rigorously reduces to the familiar expression in the ${E/{\cs EP}\rightarrow 0}$ limit.

Even without solving the relation (\ref{newpho}) for ${E'}$ in terms of ${E}$, since
${\varepsilon'< 1}$, one may be tempted to infer that at sufficiently high energies any
red-shifted photons are somewhat more red-shifted according to (\ref{newpho}) than predicted by
special relativity. This, however, would be a mistake. It is clear that no such definite
statement can be made for the case of receiver approaching toward rather than receding from the
source---i.e., for the blue-shifted photons---without actually solving the relation
(\ref{newpho}). Unfortunately, (\ref{newpho}) is hopelessly nonlinear to be solved easily. In
fact, its solution is a root of an {\it eighth} order polynomial equation in the ratio ${E'/E}$,
which, of course, according to Galois theory, does not possess any general solution in terms of
radicals. With diligence, a particular solution may be found, but instead we shall
discuss approximate solutions in the section after next, where we investigate into the
experimental verifiability of the Doppler shifts (\ref{newpho}). But first we must attend to
a theoretical and conceptual issue of utmost significance.

\section{How fast does time flow?}

It is evident that the conception of time afforded by the present theory, as encapsulated
within the fundamental line element
(\ref{quad}), is profoundly unorthodox. In particular, in addition to
motion, time now depends also on the phase space evolution of systems that record it, and hence
would be different, in general, for different recording systems. Moreover, this new conception
of time dispels, at a stroke, the spell of the `block' view of time \cite{Popper}, which is
widely thought to
be an inevitable byproduct of Einstein's special relativity. According to this `block' view,
since in the Minkowski picture time is as `laid out' {\it a priori} as space, and since space
clearly does not seem to `flow', what we perceive as a `flow of time', or `becoming', must be
an illusion. Worse still, in Einstein's theory, the relativity of simultaneous events demands
that what is `now' for one inertial observer cannot be the same, in general, for another.
Therefore, to accommodate `nows' of all possible observers, events must exist {\it a priori},
all at once, across the whole span of time \cite{Penrose-1989}. As Weyl once so aptly put it,
``The objective world simply {\it is}, it does not {\it happen}'' \cite{Weyl-1949}.
Einstein himself was quite painfully aware of this
shortcoming of his theories of relativity---namely, of their inability to capture the continual
slipping away of the present moment into the unchanging past \cite{Carnap-1963}. To be sure,
the alleged unreality of this transience of `now', as asserted by the `block' view of time, is
far from being universally accepted (see, e.g.,
\cite{Shimony-1993, Shimony-1998, Christian-2001}). However, what remains unquestionable is the
fact that there is no explicit assimilation of such a transience
in any of the established theories of fundamental physics.

By contrast, in the present theory, where proper time is defined by (\ref{quad}), the `block'
view of time endorsed by Weyl cannot be sustained. For time is now as much a `state dependent'
attribute of the world as states are time dependent attributes, and as the states of the world
do `happen' and `become', so does time. To appreciate this {\it dynamic} nature of proper time
as defined by (\ref{quad}), let us return once again to our clock that is moving and
evolving---now possibly non-uniformly---from, say, an event-state ${({\rm e}_1,\,{\rm s}_1)}$
to an event-state ${({\rm e}_5,\,{\rm s}_5)}$, in the combined space ${({\cal E},\,\xi)}$
(see FIG. 1).
According to the line element (\ref{quad}), the proper duration recorded by the clock would be
\begin{equation}
\Delta\tau=\int_{({\rm e}_1,\,{\rm s}_1)}^{({\rm e}_5,\,{\rm s}_5)}
\frac{1}{\gamma(\theta)}\;\;dt\,,\label{clock-duration}
\end{equation}
where ${\gamma(\theta)}$ is defined by (\ref{z-gamma}) as before. Now, assuming for simplicity
that the clock is not massless, we can represent its journey by the integral curve of a
timelike vector field ${V^A}$ on ${({\cal E},\,\xi)}$, defined, naturally, by
\begin{equation}
V^A\,:=\,{\cs lP}\frac{dz^A}{d\tau\;}\;{\buildrel * \over =}\;
\gamma(\theta)\left(c\,,\;{\cs lP}\theta^I\right)\label{4+2n-timelike}
\end{equation}
(${I=1,2,3,\dots,3+2n}$), such that its external components ${V^a}$
${(a=0,1,2,3)}$ would trace out, for each possible state ${s_i}$ of the clock, the
familiar four-dimensional timelike
world-lines in the Minkowski spacetime ${({\cal M},\,\eta)}$. In other words, ${V^A}$
would give rise to the usual timelike, future-directed, never vanishing, 4-velocity vector field
${V^a}$ for the clock, tangent to each of the external timelike world-lines. As a result,
the `length' of the overall enveloping world-line would be given by the proper time
(\ref{clock-duration}), whereas the `length' of the external world-line, for a given ${s_i}$,
would be given by the Einsteinian proper time
\begin{equation}
\Delta {\css tiE}=\int_{e_1}^{e_i}\frac{1}{\gamma(v)}\;dt\,,\label{yes-this}
\end{equation}
with ${\gamma(v)}$ being the usual Lorentz factor given by (\ref{x-gamma}). In FIG. 1, five
of such external timelike world-lines---one for each ${s_i}$ (i=1,2,3,4,5)---are depicted by the
blue curves with arrowheads going `upwards', and the overall enveloping world-line traced out by
${V^A}$ is
depicted by the dashed green curve going from the event-state ${({\rm e}_1,\,{\rm s}_1)}$ to the
event-state ${({\rm e}_5,\,{\rm s}_5)}$. It is at once clear from this picture that the external
world-line of the clock is not given all at once, stretched out till eternity, but `grows'
progressively further as time passes, with each temporally successive stage of the evolution of
the clock, like a tendril on a garden wall. In fact, from FIG. 1, line elements
(\ref{time1-metric}) and (\ref{e-metric}), and the condition (\ref{A-causality}),
it is easy to see that the instantaneous directional rate of this growth is simply 
\begin{equation}
U^a=\frac{V^a}{\sqrt{-\eta_{bc}V^bV^c}}\;\frac{d{\cs tE}}{d{\rm y}\;}\,,\;\;\;\;\;
{\rm with}\;\;\;\;\;\frac{d{\cs tE}}{d{\rm y}\;}\geq {\cs tP}\,,\label{rateofnow}
\end{equation}
where ${d{\rm y}:=|d{\bf y}|}$ is the
infinitesimal dimensionless phase space distance between the two successive states of the clock
(cf. definition (\ref{y-metric})), and ${d{\cs tE}}$ is the usual infinitesimal Einsteinian
proper
duration defined by (\ref{e-metric}). Thus `now' for the clock (depicted by the red dot in the
FIG. 1) moves in the future direction along its world-line, at the rate of no less than one
Planck unit of time per Planck unit of change in its physical state. Crucially, since the
4-velocity of an observer can never vanish, the lower bound on the above
rate shows that not only the `now' moves, but it {\it cannot} not move. To parody
Weyl quoted above, the objective world cannot simply {\it be}, it can only {\it happen}.

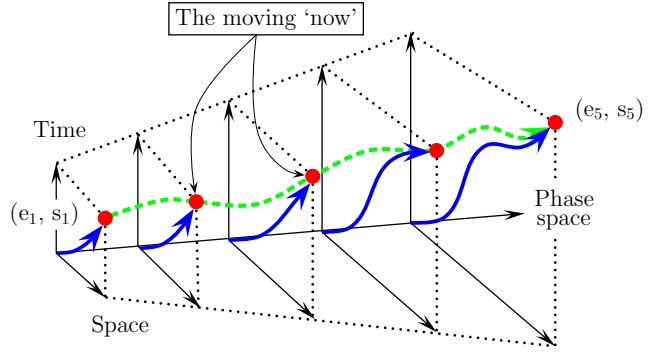
\begin{figure}
\scalebox{0.62}{
\begin{pspicture}(3.0,2.3)(14.15,9.7)

\put(12.85,5.1){\Large{\rm Phase}}

\put(12.9,4.7){\Large{\rm space}}

\put(3.35,2.4){\Large{\rm Space}}

\put(2.1,6.6){\Large{\rm Time}}

\put(5.0,9.0){\psframebox{${{}^{}}$\Large{\rm The moving `now'}${{}^{}}$}}

\psline[arrowscale=2 3]{->}(2.6,4.1)(12.65,4.95)

\psline[linewidth=0.6mm, linestyle=dotted](2.6,6.03)(3.65,4.85)

\psline[linewidth=0.6mm, linestyle=dotted](3.65,3.15)(3.65,4.85)

\psline[linewidth=0.6mm, linestyle=dotted](4.35,6.65)(5.6,5.2)

\psline[linewidth=0.6mm, linestyle=dotted](5.65,2.92)(5.6,5.2)

\psline[linewidth=0.6mm, linestyle=dotted](6.3,7.38)(8.1,5.75)

\psline[linewidth=0.6mm, linestyle=dotted](8.1,2.74)(8.1,5.75)

\psline[linewidth=0.6mm, linestyle=dotted](8.3,8.13)(10.76,6.3)

\psline[linewidth=0.6mm, linestyle=dotted](10.76,2.42)(10.76,6.3)

\psline[linewidth=0.6mm, linestyle=dotted](10.2,8.8)(13.3,6.9)

\psline[linewidth=0.6mm, linestyle=dotted](13.3,2.1)(13.3,4.4)

\psline[linewidth=0.6mm, linestyle=dotted](13.3,5.7)(13.3,6.9)

\psline[linewidth=0.6mm, linestyle=dotted](2.6,6.03)(10.25,8.82)

\psline[linewidth=0.6mm, linestyle=dotted](3.65,3.15)(13.3,2.1)

\pscurve[linewidth=0.2mm,arrowscale=2 2]{->}(6.89,8.78)(5.6,7.2)(5.58,5.37)

\pscurve[linewidth=0.2mm,arrowscale=2 2]{->}(6.89,8.78)(6.95,6.6)(7.5,5.9)
(7.93,5.77)

\psline[arrowscale=2 3]{->}(4.35,4.25)(4.35,6.65)

\psline[arrowscale=2 3]{->}(6.3,4.38)(6.3,7.38)

\psline[arrowscale=2 3]{->}(8.3,4.55)(8.3,8.13)

\psline[arrowscale=2 3]{->}(10.2,4.75)(10.2,8.85)

\psline[arrowscale=2 3]{->}(4.35,4.25)(5.7,2.92)

\psline[arrowscale=2 3]{->}(6.3,4.38)(8.15,2.68)

\psline[arrowscale=2 3]{->}(8.3,4.55)(10.76,2.4)

\psline[arrowscale=2 3]{->}(10.2,4.75)(13.3,2.1)

\pscurve[linecolor=green, linewidth=0.9mm, linestyle=dashed, arrowscale=2]{->}(3.5,4.8)
(5,5.25)(5.6,5.2)(6.9,5.15)(8.1,5.75)(9.3,6.3)(10.73,6.31)(11.8,6.8)(12.6,6.57)
(13.00,6.72)(13.2,6.8)

\psline[arrowscale=2 3]{-}(2.6,4.1)(2.6,4.6)

\psline[arrowscale=2 3]{->}(2.6,5.25)(2.6,6.03)

\psdots[linecolor=red, fillcolor=red, dotstyle=o, dotscale=2.5](3.65,4.85)(5.6,5.2)(8.1,5.75)
(10.76,6.3)(13.3,6.9)

\pscurve[linecolor=blue,linewidth=0.8mm, arrowscale=2]{->}(2.61,4.1)(3.0,4.15)(3.6,4.7)

\psline[arrowscale=2 3]{->}(2.6,4.1)(3.65,3.15)

\pscurve[linecolor=blue,linewidth=0.8mm, arrowscale=2]{->}(4.4,4.2)(4.9,4.3)(5.55,5.05)

\pscurve[linecolor=blue,linewidth=0.8mm, arrowscale=2]{->}(6.3,4.38)(7,4.48)(7.8,5.3)(8.05,5.6)

\pscurve[linecolor=blue,linewidth=0.8mm, arrowscale=2]{->}(8.3,4.55)
(8.9,4.65)(9.8,6.1)(10.627,6.27)

\pscurve[linecolor=blue,linewidth=0.8mm, arrowscale=2]{->}(10.2,4.75)
(10.9,4.85)(11.8,6.4)(12.6,6.38)(13.2,6.8)

\put(1.6,4.8){\Large ${({\rm e}_1,\,{\rm s}_1)}$}

\put(13.72,7.05){\Large ${({\rm e}_5,\,{\rm s}_5)}$}

\end{pspicture}}

\caption{Space-time-state diagram depicting the `flow of time'. (A similar diagram was used
pejoratively in \cite{Lockwood-1989} to describe and discredit Dunne's early attempt
\cite{Dunne-1927, Dunne-1938} to capture the `flow of time' in physics. The objections raised
in \cite{Lockwood-1989} and \cite{Broad-1935} against Dunne's failed attempt are
not applicable here.)}

\label{figure-1}
\end{figure}

To consolidate this conclusion, let us note that even the overall enveloping world-line (the
dashed green curve in FIG. 1) cannot remain `static' in the present scenario. This can be
seen by first parallelling the above analysis for the ${(1+2n)}$-dimensional `internal'
space ${({\cal O},\,\zeta)}$,
instead of the external spacetime ${({\cal M},\,\eta)}$. In the FIG. 1 this
amounts to slicing up the combined space ${({\cal E},\,\xi)}$ along the phase space axis
instead of the spatial axis, and then observing that even the `internal' world-line cannot but
`grow' progressively further as time passes, at the rate of
\begin{equation}
U^{\alpha}=
\frac{{\cs lP}\,V^{\alpha}}{\sqrt{-\zeta_{\beta\gamma}V^{\beta}V^{\gamma}}}\;
\frac{d{\cs tH}}{d{\rm x}\;}\,,\;\;\;\;\;
{\rm with}\;\;\;\;\;\frac{d{\cs tH}}{d{\rm x}\;}\geq c^{-1}\,,\label{rateofthen}
\end{equation}
where ${V^{\alpha}}$ is the `internal' part of the vector field ${V^A}$,
${d{\rm x}:=|d{\bf x}|}$ is the infinitesimal spatial distance between two slices, and
${d{\cs tH}}$ is the infinitesimal `internal' proper duration defined by (\ref{a-Planck}). Thus,
the `now' for the clock moves in the future direction, along its `internal' world-line, also in
the `internal' space ${({\cal O},\,\zeta)}$. Consequently, even the overall world-line---namely,
the dashed green curve in the FIG. 1---cannot be `static', but `grows' at the combined,
instantaneous, directional rate of
\begin{equation}
U^A=\left(\frac{V^a}{\sqrt{-\eta_{bc}V^bV^c}}\;\frac{d{\cs tE}}{d{\rm y}\;}\;,\,\;
\frac{{\cs lP}\,\omega^{\mu}}{\sqrt{-\zeta_{\beta\gamma}V^{\beta}V^{\gamma}}}\;
\frac{d{\cs tH}}{d{\rm x}\;}\right),\label{totalrate}
\end{equation}
where ${\omega^{\mu}\equiv V^{\mu}}$ is the instantaneous evolution rate for the clock. What is
more, this overall rate of motion for the `now' also cannot vanish. This can be seen easily by
using the rest frame for the clock to evaluate the magnitude of ${U^A}$, and then using the
lower bound from (\ref{rateofnow}) to obtain the lower bound on this magnitude, which yields
\begin{equation}
\sqrt{-\xi_{AB}U^AU^B}\;\geq\;{\cs tP}\,.\label{U-bound}
\end{equation}
Thus, in the present theory, not only does the external `now' {\it move}
along timelike world-lines,
but there does not remain even an overall `block'---such as a `static' space
${({\cal E},\,\xi)}$---that could be used to support a `block' view of time. That is to say,
the new conception of `becoming' embedded in the structure ${({\cal E},\,\xi)}$ is
{\it truly} Heraclitean \cite{Popper}.

It is also worth noting that, in the present theory, even the four-dimensional
spacetime continuum no longer enjoys the absolute status it does in Einstein's
theories of relativity. Einstein dislodged the older concepts of `absolute time' and `absolute
space', only to be replaced by the new framework of `absolute {\it spacetime}'---namely, a
continuum of {\it in principle} observable events, idealized as a connected pseudo-Riemannian
manifold with observer-independent spacetime intervals. Since it is impossible to
directly observe this remaining absolute structure without recourse to the behaviour of
material objects, it is perhaps best viewed as the `ether' of the modern times
\cite{Einstein-1920}. By contrast, it is evident from both the fundamental quadratic invariant
(\ref{quad}) and the FIG. 1 above, that in the present theory this four-dimensional spacetime
continuum no longer has the absolute, observer-independent meaning. In
fact, apart from the laws of nature, there is very little absolute structure left in the
present theory, for even the quadratic invariant (\ref{quad}) is dependent on the phase
space structure of the material system being observed. That is to say, even the manifold
${({\cal E},\,\xi)}$ that replaces the Minkowski spacetime ${({\cal M},\,\eta)}$ in the present
theory does not have the absolute status, as its construction is not independent of the system
being observed.

The presence of the absolute spacetime continuum in Einstein's theories of relativity makes them
conducive to the `block' view of time, as we discussed above. In particular, they are generally
regarded as compatible with the {\it tense-less} notions of time. To be sure, along the timelike
world-lines of material objects, the times of events in these theories are linearly ordered
{\it relative} to each other by the transitive, asymmetric, and irreflexive relation `precedes'
(see, e.g., \cite{Shimony-1993}). But there is no explicit reference to {\it absolute} past,
present, or future in this purported relation. And, in essence, it is this lack of any
reference to such tenses in the relation `precedes' along the world-lines that is responsible
for the recurring speculations on `time travel' based on these theories (see, e.g.,
\cite{Krasnikov-2002}). For, if time does not `flow' from the past to the future via the
present, then it is no different in nature from space, and hence the instants of time should be
as traversable as the places in space are. By contrast, in the theory developed here time
{\it does} `flow' from the determined past to the undetermined future, and hence `time travel'
is {\it in principle} inconceivable. Consequently, over and above its intrinsic necessity, an
experimental verification of the predictions of the present theory acquires an added incentive.
And to that empirical possibility we turn next.

\section{Experimental verifiability}

It is worth noting that, although unity in the Planck units of the quantum gravity regime, in
the ${\,m.k.s.}$ units of everyday physics the factor ${\cs tP}$ of the present theory that
converts state space distances into temporal units is some 35 orders of magnitude smaller than
the factor ${c^{-1}}$ of special relativity that converts spatial distances into temporal units.
Even in the `natural units' of the regime of particle (or quantum) physics where
${\hslash=c=1}$, the former conversion factor is some 28 orders of magnitude smaller than the
latter. And, of course, it is this minuteness of the Planck time that is responsible for the
lack of positive experimental data on physics at that scale.

In recent years, however, there have been a number of attempts to remedy this dire state of
affairs (see, e.g., \cite{Smolin-2003, Alfaro-2003, Amelino-Camelia-2003, Sarkar-2002,
Jacobson-2003} and
references therein). Most of these attempts revolve around Planck scale modifications of the
usual special relativistic dispersion relation---analogous to the one considered above
(cf. (\ref{exact-disp})). The modifying terms in these relations are usually either linearly
or quadratically suppressed by the Planck energy, and one relies on some astrophysical
phenomena---such as the gamma-ray bursts---to obtain observational bounds on the deviations from
the special relativistic predictions. As long as these deviations are linearly suppressed by
the Planck energy, astrophysical observations do tend to put useful bounds on them. However,
when the deviations happen to be quadratically suppressed by the Planck energy---as is the case
in the relation (\ref{exact-disp}) proposed above---then the minuteness of the deviations
renders even such high energy astrophysical strategy for their detection far less promising.
And yet, it has been argued in Ref.\cite{Amelino-Camelia-2003} that advanced cosmic-ray
observatories and neutrino observatories that have been planned to be operational in the near
future may provide experimental possibilities to test such quadratically suppressed
deviations (cf. also \cite{Jacobson-2003}). If this optimism of Ref. \cite{Amelino-Camelia-2003}
turns out to be justified, then the dispersion relations (\ref{exact-disp}) predicted by the
present theory may also be subjected to a test by the same means.
Since this and related possibilities are already extensively discussed in the
literature cited above, we shall not dwell on them any further. Instead, we shall concentrate
on a different possibility---that of a verifiability of the Doppler shift formula
(\ref{newpho}) predicted by the present theory.

As mentioned towards the end of Sec. VI, it is rather difficult to solve the relation
(\ref{newpho}) exactly, but it can be approximated---even at high energies---to yield a
practical result. Actually, since the experimenter is usually a receiver rather than a source
of the radiation, we shall use an equivalent formula for Doppler shifts, namely
\begin{equation}
\frac{E'}{E\;}=\frac{\sqrt{(\varepsilon')^2-\frac{v^2}{c^2}\,}}
{\,\varepsilon'\left[\,\varepsilon'+\frac{v}{c}\cos\phi'\,\right]}\,,\label{newpho-rec}
\end{equation}
written in terms of the angle ${\phi'}$---as determined by the receiver---between the relative
velocity and photon momentum. This angle is related to the corresponding angle ${\phi}$
determined by the source via the following generalized and energy-dependent aberration relation
\begin{equation}
\cos\phi'=\frac{\varepsilon'\cos\phi-\frac{v}{c}\left(\varepsilon'\right)^{-2}
+\frac{c}{v}\left(1-\left(\varepsilon'\right)^2\right)}{\varepsilon'-
\frac{v}{c}\cos\phi}\,,\label{abba-rec}
\end{equation}
which follows from the expression (\ref{aberration}) and a pair of relations such as
(\ref{angle-rela}). In the rigorous limit ${E/{\cs EP}\rightarrow 0}$ this
aberration relation reduces to the usual one derived by Einstein almost a hundred years ago.
It can be used, for instance, to obtain (\ref{newpho-rec}) from (\ref{newpho}). Of course,
(\ref{newpho-rec}) can also be easily derived directly from (\ref{source-reduced}), just as we
derived (\ref{newpho}) from (\ref{photon-reduced}) in the section before the last.

Now, a Maclaurin expansion of the right hand
side of (\ref{newpho-rec}) around the value ${E/{\cs EP}=0}$,
after keeping terms only up to the second order in the ratio ${E/{\cs EP}}$, gives
\begin{equation}
\begin{split}
\frac{\,E'}{E\,}\approx&\,\frac{\sqrt{1-\frac{v^2}{c^2}}\;\,}{1+\frac{v}{c}\cos\phi'} \\
&+\frac{1}{2}\frac{E^2}{\css E2P}\!\left[\frac{1-2\frac{v^2}{c^2}-\frac{v^3}{c^3}
\cos\phi'}{\left(1\!+\!\frac{v}{c}\cos\phi'\right)^2\sqrt{1-\frac{v^2}{c^2}}}
\right]\!\!\left(\!1-\frac{\,E'}{E\,}\right)^2\!\!\!+\dots\label{series-a} \\
\end{split}
\end{equation}
This truncation is an excellent approximation to (\ref{newpho-rec}). Even for TeV
photons, the next term in the expansion---${E^4/{\css E4P}}$---is of the order of ${10^{-64}}$.
The quadratic equation (\ref{series-a}) can now be solved for the desired ratio ${E'/E}$, and
then the physical root once again expanded, now in the powers of ${v/c}$. In what results if
we again keep terms only up to the second order in the ratio ${E/{\cs EP}}$, then, after some
straightforward algebra, we arrive at
\begin{equation}
\frac{\,E'}{E\,}\approx 1-\frac{v}{c}\cos\phi'
+\left[\left(1+\frac{1}{2}\frac{E^2}{\css E2P}\right)\cos^2\!\phi'-\frac{1}{2}\right]
\frac{v^2}{c^2}+\dots\label{prac-inter}
\end{equation}
In the limit ${E\ll{\cs EP}}$ this expansion clearly reduces to
\begin{equation}
\frac{\,E'}{E\,}\approx 1-\frac{v}{c}\cos\phi'
+\left[\cos^2\!\phi'-\frac{1}{2}\right]\frac{v^2}{c^2}+\dots,\label{newpa-rela}
\end{equation}
which is simply the familiar special relativistic result.

Comparing (\ref{prac-inter}) and (\ref{newpa-rela}) we see that up to the first order in ${v/c}$
there is no difference between the special relativistic result and that of the present theory.
The first deviation between the two theories occur in the second-order coefficient, precisely
where special relativity differs also from the classical theory. What is more, this second-order
deviation depends non-trivially on the angle between the relative velocity and photon
momentum. For instance, up to the second order, both red-shifts (${\phi'=0}$) and blue-shifts
(${\phi'=\pi}$) predicted by (\ref{prac-inter}) differ significantly from those predicted by
special relativity. In particular, the red-shifts are now somewhat less red-shifted, whereas
the blue-shifts are somewhat more blue-shifted. On the other hand, the transverse red-shifts
(${\phi'=\pi/2}$ or ${3\pi/2}$) remain identical to those predicted by special relativity. As
a result, even for the photon with energy approaching the Planck energy, an Ives-Stilwell type
classic experiment \cite{Ives-Stilwell} would not be able to distinguish the predictions of
the present theory from those of special relativity. The complete angular distribution of the
second-order coefficient predicted by the two theories, along with its energy dependence, is
displayed in the FIG. 2.

In spite of this non-trivial angular dependence of Doppler shifts, in practice, due to the
quadratic Planck energy suppression, distinguishing the expansion (\ref{prac-inter}) from its
special relativistic counterpart
(\ref{newpa-rela}) would be a highly non-trivial task. This can also be inferred from FIG. 2,
which makes it clearer that a near-future verification of the prediction (\ref{prac-inter}) by
means of a terrestrial experiment is highly unlikely. The maximum laboratory energy that may
be available to us is of the order of TeV, yielding ${E^2/{\css E2P}\sim 10^{-32}}$. This
represents a correction of one part in ${10^{32}}$ from (\ref{newpa-rela}), demanding a
phenomenal sensitivity for its detection, well beyond the means of the
state-of-the-art precision technology.

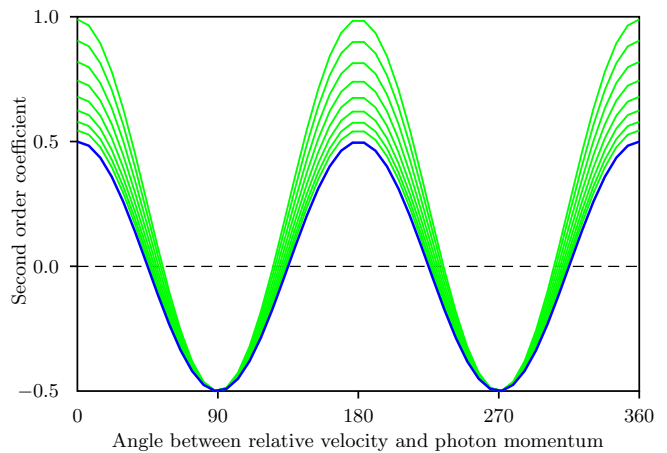
\begin{figure}
\scalebox{.83}{
\begin{pspicture}(-0.5,-2.7)(8.87,4.3)
\psset{xunit=.25mm,yunit=4cm}
\psaxes[axesstyle=frame,tickstyle=bottom,
dx=90\psxunit,Dx=90,Ox=0,dy=0.5\psyunit,Dy=0.5,Oy=-0.5](0,-0.5)(360,1)
\psplot[linewidth=.2mm,linestyle=dashed]{0.0}{255}{x 0.0}
\psplot[linewidth=.3mm,
linecolor=green]{0.0}{360}{x dup cos exch cos mul 1.490 mul 0.5 neg add}
\psplot[linewidth=.3mm,
linecolor=green]{0.0}{360}{x dup cos exch cos mul 1.405 mul 0.5 neg add}
\psplot[linewidth=.3mm,
linecolor=green]{0.0}{360}{x dup cos exch cos mul 1.320 mul 0.5 neg add}
\psplot[linewidth=.3mm,
linecolor=green]{0.0}{360}{x dup cos exch cos mul 1.245 mul 0.5 neg add}
\psplot[linewidth=.3mm,
linecolor=green]{0.0}{360}{x dup cos exch cos mul 1.180 mul 0.5 neg add}
\psplot[linewidth=.3mm,
linecolor=green]{0.0}{360}{x dup cos exch cos mul 1.125 mul 0.5 neg add}
\psplot[linewidth=.3mm,
linecolor=green]{0.0}{360}{x dup cos exch cos mul 1.080 mul 0.5 neg add}
\psplot[linewidth=.3mm,
linecolor=green]{0.0}{360}{x dup cos exch cos mul 1.045 mul 0.5 neg add}
\psplot[linewidth=.4mm,linecolor=blue]{0.0}{360}{x dup cos exch cos mul 0.5 neg add}
\put(0.55,-2.9){Angle between relative velocity and photon momentum}
\put(-1.05,-0.63){\rotatebox{90}{Second order coefficient}}
\end{pspicture}}
\caption{The energy-dependent signatures of refined relativity. The green curves are based on
the predictions of the present theory, for ${E/{\cs EP}=0.3}$ to ${0.99}$ in the ascending
order, whereas the blue curve is based on the prediction of special relativity.}
\label{figure-2}
\end{figure}

Let us see if we can do better than this by using an extraterrestrial source. For example,
suppose we attempt to distinguish between the second order Doppler shifts predicted by the two
theories using a binary pulsar emitting ${\gamma}$-ray pulses. Of course, most of the known
pulsars are radio sources, but a handful of them are indeed ${\gamma}$-ray pulsars emitting
photons of energy in the range of TeV (see, e.g., \cite{Atoyan-2002}). Now, it is well known
that binary pulsars not only exhibit Doppler shifts, but the second-order shifts resulting from
the rapid motion of a pulsar orbiting about its companion can be isolated, say, from the first
order shifts, because they depend on the square of the relative velocity, which varies as the
pulsar moves
along its two-body elliptical orbit (see, e.g., \cite{Will-1993}). Due to these Doppler shifts,
the rate at which its pulses are observed on Earth reduces slightly when the pulsar is receding
from the Earth in its orbit (${\phi'=0}$), compared to when it is approaching the Earth
(${\phi'=\pi}$). The parameter relevant in the `time of arrival' analysis of these pulses is
variously called the `Red-shift-Doppler parameter' or the `time dilation parameter'
\cite{Will-1993}, which is a non-trivial function of the gravitational red-shift, the masses of
the two binary stars, and other Keplerian parameters. For a radio pulsar also exhibiting
periastron precession similar to the perihelion advance of Mercury, this parameter can be
determined with an excellent precision. This is, of course, the case for the most famous
pulsar: PSR
1913+16 \cite{Will-2001}. Now, the arrival times of the pulses of this pulsar---which have been
monitored for three decades---are extremely sensitive to the `time dilation' parameter,
and thereby to the second-order Doppler shifts brought about by its orbital motion.
What is more, the overall precision on the timing of the pulses of PSR 1913+16, and consequently
that of its periastron advance, is famously better than one part in ${10^{12}}$. Indeed, the
monitoring of the decaying
orbit of PSR 1913+16 constitutes one of the most precise tests of general relativity to date.

Encouraged by these facts, one might hope that---at least in principle---similar careful
observations of a suitable ${\gamma}$-ray pulsar may be able to distinguish the predictions of
the present theory from those of special relativity. However, the highest energy radiating
pulsar known to date emits ${\gamma}$-ray photons of energy no greater than 10 TeV, giving the
discriminating ratio ${E^2/{\css E2P}}$ to be of the order of ${10^{-30}}$, which is only two
orders of magnitude improvement from a possible terrestrial scenario. Thus, it appears that,
even with such an exotic astrophysical source as a ${\gamma}$-ray pulsar, it would be quite a
challenge to distinguish between the predictions of the two theories.

Facing this difficulty, one may ask a converse question: How energetic the ${\gamma}$-rays
emitted by a pulsar have to be to meet the achievable precision? Even if we dare to go by the
remarkable precision available on PSR 1913+16, the answer would have to be: exceeding
${10^{10}}$ TeV.

Of course, no such ultra-high energy pulsar has been found. Worse still, there are reasons
to believe---at least within the standard framework of special relativity---that none
above 10 TeV can ever be found. Above the 10 TeV threshold, ${\gamma}$-rays are expected to
severely attenuate through pair-production in the intervening infrared background long
before reaching the Earth. There are, however, good indications that this theoretical threshold
is, in fact, not respected by nature \cite{Aharonian-1999, Finkbeiner-2000}. Perhaps within a
refined inertial structure, such as the one being proposed here, there is no such threshold
(cf. \cite{Sarkar-2002, Alfaro-2003}). Whatever turns out to be the resolution of this conflict,
there may be independent reasons for a pulsar to refrain from emitting radiations of energy as
high as ${10^{10}}$ TeV. For instance, the intrinsic dynamics of the magnetically trapped
charged particles responsible for producing the radiation emitted by the pulsars may be subject
to its very own high energy threshold. We simply do not know. On balance, however, we cannot
rule out the possibility of finding in the future a suitable ${\gamma}$-ray binary pulsar
emitting radiation of energies exceeding ${10^{10}}$ TeV.

The discussion above is clearly meant to be indicative of the difficulties rather than that of
realistic possibilities. And even then we face almost insurmountable obstacles to our aim.
This is a great pity, not the least because an experimental test of the present theory has a
direct bearing on one of the oldest disputes in natural philosophy: Does the perceived `flow of
time' reflect a genuinely structural attribute of the world, as the present theory maintains,
or is it an illusion, as special relativity seems to suggest? In view of the issues discussed
in the previous section, an empirical corroboration of a signature predicted by the present
theory would lend much needed support to the former---Heraclitean---notion of time.

\section{CONCLUDING REMARKS}

Perhaps one of the most attractive features of the above theory is its economy of thought,
parallelling the rationale of Einstein's special relativity. In particular, unlike the
two-scales theories mentioned in the Introduction, the present theory is based on only one
observer-independent fundamental scale, namely, the inverse of the Planck time, and this fact
makes it a truly Planck scale rooted theory. To be sure, the vacuum speed of light also remains
an invariant in the theory, but it plays only a secondary and derivative role. And yet, this
generalization of Einstein's second postulate does not necessitate any compromise with his
first postulate, namely, the principle of relativity. Quite the contrary, a more complete
implementation of this basic principle offers at least four-fold improvement over his special
relativity. First, it allows one to eliminate the dualistic notion of time pervasive in our
physical theories---from classical to quantal---that has caused so many conceptual difficulties
in our understanding of `quantum gravity'. Second, it allows one to capture and quantify the
elusive `flow of time' as a genuinely structural attribute of the world. Third, it renders
redundant one of the last remnants of `ether'---namely, the absolute spacetime---from the local
inertial structure. And fourth, it allows one to eliminate the unphysical concepts of unbounded
energies, momenta, lengths, and durations from a physical theory.

These are more than sufficient reasons to take the present theory seriously, despite its
unorthodox appeal. But the theory has even more to offer. For example, it provides a natural
contraception for the speculations on time travel. A prerequisite for the possibility of time
travel is a tenseless structure of spacetime, necessitated by the special and general
theories of relativity. By contrast, the notions of past, present, and future are intrinsic
to the spacetime structure proposed in the present theory, and hence time travel becomes
{\it in principle} inconceivable.

Another somewhat related implication of the present theory
concerns the thermodynamic arrow of time. It is well known that---from Boltzmann to
Prigogine---no one has succeeded in explaining the thermodynamic unidirectionality of time in
terms of micro-physics, essentially because the microscopic laws of physics happen to be
symmetric in time. But according to the present theory the local causal structure intrinsically
distinguishes the future from the past, due to the built-in directional `flow of time', and
thereby provides an opportunity to derive the thermodynamic arrow of time
from the time-symmetric microscopic laws. This aspect of the proposed theory has not been
discussed in the present paper, but it will be explicated fully in a separate publication.

In addition to these theoretical implications, several phenomenological aspects of the present
theory are also worth noting. As mentioned before, several attempts to construct a theory of
quantum gravity predict energy-dependent deviations from the dispersion relations based on
special relativity. The present theory shows that such deviations can be understood on the
basis of the first principles of a theory, thereby providing a natural physical
understanding of the deviations. The prediction of these deviations, along with the deviations
from the special relativistic Doppler shifts, suggest that the present theory may be viewed
also as a non-artificial test-theory for the experimental investigations of special relativity. 

Given the refined local inertial structure embedded in the quadratic invariant (\ref{quad}),
the next natural challenge, of course, is to understand how this new structure refines the
classical---and, indeed, the quantum---conceptions of gravity as prescribed by the principle
of equivalence. This challenge, with quite a broad understanding of the term `quantum gravity',
will be taken up and addressed fully in a companion paper.

\vfill\eject

\begin{acknowledgments}

I am deeply indebted to my mentor Abner Shimony for his kind, generous, and
sustained financial support without which this work would not have been possible.

\end{acknowledgments}

\end{document}